\documentclass[11pt]{article}
\usepackage{amsmath}
\usepackage{amsfonts}
\usepackage{amssymb}
\usepackage[dvips]{epsfig}

 \advance \textwidth by -2in
 \advance \textheight by -3in
\def\##1{\underline{#1}}
\def\=#1{\underline{\underline{#1}}}

\def\+#1{\underline{\bf #1}}
\def\*#1{\underline{\underline{\bf #1}}}

\def\eps{\epsilon}
\def\epso{\epsilon_{\scriptscriptstyle 0}}
\def\muo{\mu_{\scriptscriptstyle 0}}
\def\ko{k_{\scriptscriptstyle 0}}
\def\co{c_{\scriptscriptstyle 0}}

\def\.{\mbox{ \tiny{$^\bullet$} }}

\def\le{\left(}
\def\ri{\right)}
\def\les{\left[}
\def\ris{\right]}
\def\lec{\left\{}
\def\ric{\right\}}

\def\l#1{\label{#1}}
\def\r#1{(\ref{#1})}

\begin{document}

\begin{center} {\bf {\LARGE On electromagnetics of an isotropic chiral medium
moving at constant velocity}} \end{center} \vskip 0.2cm

\noindent  {\bf Tom G. Mackay$^a$ and Akhlesh Lakhtakia$^b$}
\vskip 0.2cm

\noindent {\sf $^a$ School of Mathematics\\
\noindent James Clerk Maxwell Building\\
\noindent University of Edinburgh\\
\noindent Edinburgh EH9 3JZ, United Kingdom\\
email: T.Mackay@ed.ac.uk} \vskip 0.4cm

\noindent {\sf $^b$ CATMAS~---~Computational \& Theoretical Materials Sciences Group \\
\noindent Department of Engineering Science \& Mechanics\\
\noindent 212 Earth \& Engineering Sciences Building\\
\noindent Pennsylvania State University, University Park, PA
16802--6812\\
email: akhlesh@psu.edu} \vskip 0.4cm

\begin{center} {\bf Abstract} \end{center}

A medium which is an isotropic chiral medium from the perspective
of a co--moving observer is a Faraday chiral medium (FCM) from the
perspective of a non--co--moving observer. The Tellegen
constitutive relations for this FCM are established. By an
extension of the Beltrami field concept, these constitutive
relations are exploited to show that planewave propagation is
characterized by four generally independent wavenumbers. This FCM
can support negative phase velocity  at certain translational
velocities and with certain wavevectors, even though the
corresponding isotropic chiral medium does not. The  constitutive
relations and Beltrami--like fields are also used to develop a
convenient spectral representation of the dyadic Green functions
 for the FCM.

\vskip 0.2cm \noindent {\bf Keywords:} {\em Beltrami field, Bohren transform,
dyadic Green function, Faraday chiral medium, isotropic chiral medium,
Lorentz transformation, negative phase
velocity  }

\vskip 0.4cm

\newpage

\section{Introduction}

A fundamental issue in electromagnetics is  the variation in the
perceived properties of a linear medium according to the observer's
inertial frame of reference. Interest in this topic dates from  the
earliest days of the special theory of relativity and it remains an
active area of research. Electromagnetic fields in mediums which are
isotropic dielectric--magnetic from the perspective of a co--moving
observer  have been widely studied (Chen 1983; Kong 1986; Pappas
1965).  Recent studies involving an isotropic dielectric--magnetic
medium have demonstrated that planewave propagation with
 positive, negative or orthogonal phase velocity can arise
 depending upon the observer's inertial frame of reference (Mackay \& Lakhtakia 2004a;
  Mackay \emph{et al.} 2006).
The electromagnetics of simply moving  plasmas have also been extensively
investigated (Chawla \& Unz 1971).

In this paper we consider electromagnetic fields in linear
isotropic chiral mediums moving at constant velocities.  A natural
formalism for investigating the electromagnetic properties of an
isotropic chiral medium, as observed from a co--moving inertial
frame of reference,  is provided by Beltrami fields. The defining
characteristic of a Beltrami field is that the  curl of the
field is a
 scalar multiple of the field itself (Lakhtakia 1994a,b). These fields are useful
for analysis of  a wide range physical phenomenons,  as in
astrophysics (Chandrasekhar 1956, 1957),
 hydrodynamics and magnetohydrodynamics (Dritschel
1991), thermoacoustics (Ceperley 1992), chaotic flows (McLaughlin
and Pironneau 1991), plasma physics (Yoshida 1991) and
magnetostatics (Marcinkowski 1992), for example. In the following
sections, an extension of the Beltrami field concept is developed
to investigate the electromagnetic properties of an isotropic
chiral medium as observed from a non--co--moving inertial frame of reference.

In earlier studies involving  isotropic chiral mediums moving at a
constant velocity, the Lorentz--transformed wavevector and
Lorentz--transformed angular frequency have been utilized to
explore Doppler shift and aberration (Engheta \emph{et al.} 1989);
the scattering response of an electrically small sphere made of an
isotropic chiral medium   has been formulated (Lakhtakia \emph{et
al.} 1991); and planewave propagation   has been investigated for
relatively low translational speeds (Hillion 1993; Ben--Shimol \&
Censor 1995, 1997). Reflection and transmission coefficients for a
uniformly moving isotropic chiral slab have also been calculated
using the Lorentz--transformed electromagnetic fields (Hinders
\emph{et al.} 1991).

In contrast to these earlier works, the analysis presented in the
following sections begins with a derivation of the Tellegen
constitutive relations, from the perspective of a non--co--moving
observer, for a medium which is an isotropic chiral medium for a
co--moving observer. By means of the Bohren transform and the
consequent introduction of Beltrami--like fields, these
 constitutive relations are
exploited to consider planewave propagation~---~specifically, the
propensity for negative phase velocity~---~from the perspective of
a non--co--moving observer. The  constitutive relations, combined
with  Beltrami--like fields, are also used to develop a convenient
spectral representation of the dyadic Green functions
 for the isotropic chiral medium moving at constant velocity.

As regards notational matters, 3--vectors are underlined whereas 3$\times$3 dyadics are double underlined.
The identity 3$\times$3 dyadic is written as $\=I$.
 Vectors with the $\hat{}$ symbol overhead are unit vectors. The operators $\mbox{Re} \lec \. \ric$ and
$\mbox{Im} \lec \. \ric$ deliver the real and imaginary parts,
respectively,  of complex--valued quantities; the superscript
${}^*$ denotes a complex conjugate; and $i = \sqrt{-1}$. The
permittivity and permeability of free space are denoted $\epso$
and $\muo$, respectively;
  $\co=1/\sqrt{\epso\muo}$
is the speed of light in free space.

\section{Constitutive relations} \l{CR_section}

We contrast the electromagnetic properties in
 two different inertial frames of reference, denoted as $\Sigma'$
and $\Sigma$, where $\Sigma'$ moves with constant velocity
$\#v=v\hat{\#v}$ relative to $\Sigma$. The spacetime coordinates
$\le \#r', t' \ri$ in frame $\Sigma'$ are related to the spacetime
coordinates $\le \#r, t \ri$ in frame $\Sigma$ by the Lorentz
transformation
\begin{equation}
\#r' =  \=Y\.\#r - \gamma \#v t\,,\qquad
 t' = \displaystyle{\gamma \le t - \frac{\#r\.\#v}{\co^2} \ri}\,,
\end{equation}
wherein
\begin{equation}
\=Y = \=I + \le \gamma - 1 \ri \hat{\#v} \, \hat{\#v}\,,
\qquad \gamma = \frac{1}{\sqrt{1-\beta^2}}\,,
\end{equation}
and the relative speed $\beta = v / \co$.

In the absence of
sources, the spatiotemporal variations of the (time--domain) electromagnetic fields in frame
$\Sigma'$ are related as
\begin{equation}
\left.
\begin{array}{l}
\displaystyle{
\nabla' \times \#{\check E}' (\#r', t') + \frac{\partial}{\partial t'}   \#{\check B}' (\#r', t') = \#0} \\
\vspace{-3mm}\\
\displaystyle{\nabla' \times \#{\check H}'(\#r', t') -
\frac{\partial}{\partial t'} \#{\check D}'(\#r', t') = \#0}
\end{array}
\right\}, \l{Sigma_p}
\end{equation}
whereas those in frame $\Sigma$ are related as
\begin{equation}
\left.
\begin{array}{l}
\displaystyle{\nabla \times \#{\check E}(\#r, t) + \frac{\partial}{\partial t} \#{\check B} (\#r, t) = \#0} \\
\vspace{-3mm}\\
\displaystyle{\nabla \times \#{\check H}(\#r, t) -
\frac{\partial}{\partial t} \#{\check D} (\#r, t) = \#0}
\end{array}
\right\},  \l{Sigma}
\end{equation}
as dictated by the Lorentz covariance of the Maxwell curl postulates.
The primed and unprimed fields in \r{Sigma_p} and \r{Sigma} are
connected via
 the Lorentz transformation as (Chen 1983)
\begin{eqnarray}
\#{\check E}' (\#r', t') &=& \gamma \,\les \=Y^{-1} \.\#{\check E}
(\#r,t) + \#v \times
\#{\check B} (\#r,t) \ris , \l{Ep}  \\
\#{\check B}' (\#r',t') &=& \gamma \, \les\=Y^{-1} \. \#{\check
B}(\#r,t) - \frac{1}{\co^2} \,\#v \times
\#{\check E} (\#r,t) \ris ,  \l{Bp}\\
\#{\check H}' (\#r',t') &=& \gamma \, \les \=Y^{-1} \. \#{\check
H} (\#r,t)- \#v \times
\#{\check D}(\#r,t)\ris , \l{Hp} \\
\#{\check D}' (\#r',t') &=& \gamma \,\les\=Y^{-1} \. \#{\check D}
(\#r,t)+ \frac{1}{\co^2} \,\#v \times \#{\check H} (\#r,t) \ris.
\l{Dp}
\end{eqnarray}

Let us consider  a homogeneous medium, which is an  isotropic chiral
medium from the perspective of an observer co--moving relative to
the frame $\Sigma'$. In the most general scenario, the medium is
spatiotemporally nonlocal.
 From the perspective of the observer co--moving relative to $\Sigma'$, the constitutive
relations of the medium may be expressed in the Tellegen form as
(Lakhtakia 1994b)
\begin{eqnarray}
\#{\check D}'(\#r',t') &=&   \epso \int_{\#s'} \int_{u'}
\check{\eps}' (\#s',
u')\, \#{\check E}' (\#r' - \#s',t'-u') \; du' d \#s' \nonumber \\
&& + i \sqrt{\epso \muo} \int_{\#s'} \int_{u'} \check{\xi}' (\#s',
u')\,
 \#{\check H}'(\#r' - \#s',t'-u')\; du' d\#s',\l{Con_rel_p1} \\
 \#{\check B}'(\#r',t') &=&  - i \sqrt{\epso \muo}  \int_{\#s'}
 \int_{u'} \check{\xi}'
 (\#s', u')\,
\#{\check E}'(\#r' - \#s',t'-u') \; du' d\#s' \nonumber \\ && + \muo
\int_{\#s'} \int_{u'} \check{\mu}' (\#s', u')\, \#{\check H}' (\#r'
- \#s',t'-u') \; du' d\#s' , \l{Con_rel_p2}
\end{eqnarray}
with the real--valued (time--domain) constitutive parameters
$\check{\eps}' (\#r', t')$, $\check{\xi}' (\#r', t')$ and
$\check{\mu}' (\#r', t')$. By implementing the spatiotemporal
Fourier transforms  (Lakhtakia \& Weiglhofer 1996)
\begin{equation}
 \displaystyle{ \lec \begin{array}{c} \#E'_\sharp
(\#k', \omega') \vspace{2mm} \\
\#H'_\sharp (\#k', \omega') \vspace{2mm} \\
\#D'_\sharp (\#k', \omega') \vspace{2mm} \\ \#B'_\sharp (\#k',
\omega')
\end{array} \ric = \int_{\#r'} \int_{t'} \lec \begin{array}{c}
\#{\check E}' (\#r',
t') \vspace{2mm} \\
\#{\check H}' (\#r', t')
\vspace{2mm} \\
\#{\check D}' (\#r', t') \vspace{2mm} \\ \#{\check B}' (\#r', t')
 \end{array} \ric  \exp
\les i \le \omega' t' - \#k' \. \#r' \ri \ris \; dt' d\#r' }
\l{FT_1}
\end{equation}
and
\begin{equation}
 \displaystyle{\chi'_\sharp (\#k', \omega') =
\int_{\#r'} \int_{t'} \check{\chi}' (\#r', t') \exp \les i \le
\omega' t' - \#k' \. \#r' \ri \ris \; dt' d\#r', \hspace{15mm} \le
\chi = \eps, \xi, \mu \ri }, \l{FT_2}
\end{equation}
along with the convolution theorem (Walker 1988),
the frequency--domain constitutive relations emerge as
\begin{equation}
\left.
\begin{array}{l}
 \#D'_\sharp (\#k', \omega') = \epso\, \eps'_\sharp (\#k', \omega' ) \,\#E'_\sharp (\#k', \omega' ) +
 i \sqrt{\epso \muo} \, \xi'_\sharp (\#k', \omega' )
  \,
 \#H'_\sharp(\#k', \omega')
\\ \vspace{-3mm} \\
\#B'_\sharp (\#k', \omega') = - i \sqrt{\epso \muo} \,\xi'_\sharp
(\#k', \omega' ) \, \#E'_\sharp (\#k', \omega') + \muo\,
\mu'_\sharp (\#k', \omega' ) \,\#H'_\sharp (\#k', \omega')
\end{array}
\label{cr_FD_1} \right\}.
\end{equation}

In many applications the effects of spatial nonlocality are
negligible in comparison to those of temporal nonlocality.
The  constitutive relations \r{cr_FD_1}
 may then be approximated as
\begin{equation}
\left.
\begin{array}{l}
 \#D' (\#r', \omega') = \epso\, \eps'_{\,\mbox{\tiny{}}} ( \omega' ) \,\#E' (\#r', \omega' ) +
 i \sqrt{\epso \muo} \, \xi'_{\,\mbox{\tiny{}}} ( \omega' )
  \,
 \#H'(\#r', \omega')
\\ \vspace{-3mm} \\
\#B' (\#r', \omega') = - i \sqrt{\epso \muo}
\,\xi'_{\,\mbox{\tiny{}}} ( \omega' ) \, \#E' (\#r', \omega') +
\muo\, \mu'_{\,\mbox{\tiny{}}} ( \omega' ) \,\#H' (\#r', \omega')
\end{array}
\label{cr_FD_2} \right\},
\end{equation}
for spatially local mediums, wherein
\begin{equation}
 \displaystyle{ \lec \begin{array}{c} \#E'
(\#r', \omega') \vspace{0mm} \\
\#H' (\#r', \omega') \vspace{0mm} \\
\#D' (\#r', \omega') \vspace{0mm} \\ \#B' (\#r', \omega')
\end{array} \ric =  \int_{t'} \lec \begin{array}{c}
\#{\check E}' (\#r', t') \vspace{
0 mm} \\
\#{\check H}' (\#r', t')
\vspace{0mm} \\
\#{\check D}' (\#r', t') \vspace{0mm} \\ \#{\check B}' (\#r', t')
 \end{array} \ric  \exp
\le i  \omega' t' \ri \; dt'  } \l{FT_1x}
\end{equation}
and
\begin{equation}
 \displaystyle{\chi' ( \omega') =
\int_{t'} \check{\check{\chi}}'  ( t') \exp \le i  \omega' t'
 \ri  \; dt' , \hspace{15mm} \le \chi = \eps, \xi,
\mu \ri }, \l{FT_2x}
\end{equation}
with $ \check{\check{\chi}}'  (t') \equiv  \check{\chi}' (\#r',
t')$ for $\chi = \eps$, $\xi$ and $\mu$.

Let us now proceed to develop the frequency--domain constitutive
relations in the reference frame $\Sigma$. After using \r{Ep}, \r{Hp}
and \r{Dp} to substitute for $\#{\check E}' (\#r', t')$,
$\#{\check H}' (\#r', t')$ and $\#{\check D}' (\#r', t')$,
respectively,
 the constitutive relation \r{Con_rel_p1}
may be expressed in terms of $\Sigma$ fields as
\begin{eqnarray}
  \#{\check D}(\#r,t) &=& \epso \int_{\#s'} \int_{u'}\Bigg(
 \check{\eps}_{}' (\#s', u') \#{\check E}(\#r
- \=Y\.\#s' - \gamma \#v u', t - u'\gamma  - \frac{\gamma
\#s'\.\#v}{\co^2} )  \nonumber
\\ && + \=Y\. \lec \#v \times \les
\check{\eps}_{}' (\#s', u')
 \#{\check B}(\#r - \=Y\.\#s' - \gamma \#v u', t - u'\gamma  -
\frac{\gamma \#s'\.\#v}{\co^2} )
  \ris \ric \Bigg) \; du' d\#s' \nonumber \\
&& + i \sqrt{\epso \muo} \int_{\#s'} \int_{u'} \Bigg(
\check{\xi}_{}' (\#s', u') \#{\check H}(\#r - \=Y\.\#s' - \gamma
\#v u', t - u'\gamma  - \frac{\gamma \#s'\.\#v}{\co^2} )\nonumber
\\ &&
- \=Y\.\lec \#v \times \les  \check{\xi}_{}' (\#s', u') \#{\check
D}(\#r - \=Y\.\#s' - \gamma \#v u', t - u'\gamma  - \frac{\gamma
\#s'\.\#v}{\co^2} ) \ric \ris\Bigg) \; du' d\#s' \nonumber
\\&&   - \=Y\. \les  \frac{1}{\co^2} \#v
\times \#{\check H}(\#r,t)\ris. \l{ee1}
\end{eqnarray}
Similarly, the constitutive relation \r{Con_rel_p2} may be
expressed in terms of $\Sigma$ fields as
\begin{eqnarray}
 \#{\check B}(\#r,t) &=& -i \sqrt{\epso \muo} \int_{\#s'} \int_{u'} \Bigg(
 \check{\xi}_{}' (\#s', u') \#{\check E}(\#r
- \=Y\.\#s' - \gamma \#v u', t - u'\gamma  - \frac{\gamma
\#s'\.\#v}{\co^2} ) \nonumber
\\ && + \=Y\.\lec \#v \times  \les
\check{\xi}_{}' (\#s', u')
 \#{\check B}(\#r - \=Y\.\#s' - \gamma \#v u', t - u'\gamma  -
\frac{\gamma \#s'\.\#v}{\co^2} )
 \ris \ric \Bigg) \; du' d\#s' \nonumber \\
&& + \muo  \int_{\#s'} \int_{u'} \Bigg(  \check{\mu}_{}' (\#s',
u') \#{\check H}(\#r - \=Y\.\#s' - \gamma \#v u', t - u'\gamma  -
\frac{\gamma \#s'\.\#v}{\co^2} ) \nonumber \\
&& - \=Y\. \lec  \#v \times \les \check{\mu}_{}' (\#s', u')
\#{\check D}(\#r - \=Y\.\#s' - \gamma \#v u', t - u'\gamma  -
\frac{\gamma \#s'\.\#v}{\co^2} ) \ris \ric \Bigg) \; du' d\#s'
\nonumber
\\&&   + \=Y \les  \frac{1}{\co^2} \#v
\times \#{\check E}(\#r,t) \ris, \l{mm1}
\end{eqnarray}
by using \r{Ep}, \r{Hp} and \r{Bp} to substitute for $\#{\check E}'
(\#r', t')$, $\#{\check H}' (\#r', t')$ and $\#{\check B}' (\#r',
t')$, respectively. Implementation of the spatiotemporal Fourier
transforms
\begin{equation}
 \displaystyle{ \lec \begin{array}{c} \#E_\sharp
(\#k, \omega) \\
\#H_\sharp (\#k, \omega) \\
\#D_\sharp (\#k, \omega) \\ \#B_\sharp (\#k, \omega) \end{array}
\ric = \int_{\#r} \int_{t} \lec \begin{array}{c} \#{\check E} (\#r,
t) \\
\#{\check H} (\#r, t)
 \\
\#{\check D} (\#r, t)
 \\ \#{\check B} (\#r,
t)
 \end{array} \ric  \exp
\les i \le \omega t - \#k \. \#r \ri \ris \; dt\, d\#r } \l{FT_3}
\end{equation}
and \r{FT_2} with \r{ee1} delivers
\begin{eqnarray}
 \#D_\sharp (\#k,\omega) &=& \epso \eps'_\sharp (\#k' , \omega')  \lec    \#E_\sharp
(\#k,\omega)
  + \=Y \. \les \#v \times
 \#B_\sharp (\#k,\omega)
 \ris  \ric \nonumber \\
&& + i \sqrt{\epso \muo}\, \xi'_\sharp (\#k' , \omega') \lec
\#H_\sharp (\#k,\omega)
  - \=Y\. \les \#v \times
\#D_\sharp (\#k,\omega)  \ris \ric \nonumber
\\&&  - \=Y \. \les \frac{1}{\co^2} \#v
\times \#H_\sharp (\#k,\omega)\ris, \l{ee2}
\end{eqnarray}
and with  \r{mm1} yields
\begin {eqnarray}
 \#B_\sharp (\#k,\omega) &=& -i \sqrt{\epso \muo} \, \xi'_\sharp (\#k', \omega' ) \lec
\#E_\sharp (\#k,\omega)    + \=Y \. \les  \#v \times
  \#B_\sharp (\#k,\omega)
 \ris \ric \nonumber \\
&& + \muo \mu'_\sharp (\#k', \omega' ) \lec  \#H_\sharp
(\#k,\omega)  - \=Y \. \les  \#v \times  \#D_\sharp (\#k,\omega)
 \ris\ric \nonumber
\\&&   + \=Y \. \les  \frac{1}{\co^2} \#v
\times \#E_\sharp (\#k,\omega) \ris. \l{mm2}
\end{eqnarray}
In the derivation of \r{ee2} and \r{mm2}, the principle of
phase invariance (Pappas 1965; Kong 1986) has been invoked to obtain the relations
\begin{equation}
\displaystyle{\#k' =  \=Y\.\#k  - \frac{\omega \gamma}{\co^2}\, \#v}\,,\qquad
 \omega' = \gamma \, \le \omega - \#k\. \#v \ri
\,. \l{phase_invar}
\end{equation}

 The two independent expressions \r{ee2} and \r{mm2} which relate
$\#D_\sharp (\#k, \omega)$ and $\#B_\sharp (\#k, \omega)$ to
$\#E_\sharp (\#k, \omega)$ and $\#H_\sharp (\#k, \omega)$ can be
manipulated to deliver the $\Sigma$ frequency--domain constitutive
relations
\begin{equation}
\left.
\begin{array}{l}
 \#D_\sharp (\#k, \omega) = \epso\, \=\eps_\sharp (\#k', \omega' ) \.\#E_\sharp (\#k, \omega ) +
 i \sqrt{\epso \muo} \, \=\xi_\sharp (\#k', \omega' )
  \.
 \#H_\sharp (\#k, \omega)
\\ \vspace{-3mm} \\
\#B_\sharp (\#k, \omega) = - i \sqrt{\epso \muo} \, \=\xi_\sharp
(\#k', \omega' ) \. \#E_\sharp (\#k, \omega) + \muo\, \=\mu_\sharp
(\#k', \omega' ) \.\#H_\sharp (\#k, \omega)
\end{array}
\label{cr1} \right\}.
\end{equation}
Herein the 3$\times$3 constitutive dyadics all have the form
\begin{equation}
 \=\chi_\sharp (\#k', \omega' ) = \chi^t_\sharp (\#k', \omega' ) \, \=I - i
\chi^{g}_\sharp (\#k', \omega' ) \hat{\#v} \times \=I + \les
\chi^{z}_\sharp (\#k', \omega' ) - \chi^t_\sharp (\#k', \omega' )
\ris \hat{\#v} \, \hat{\#v},  \hspace{8mm} (\chi = \eps, \xi,
\mu).\l{chi_dyadic}
\end{equation}
The unprimed constitutive parameters emerge as
\begin{eqnarray}
&& \left. \begin{array}{l} \displaystyle{ \eps^t_\sharp (\#k',
\omega' ) = \eps'_\sharp (\#k', \omega' ) \lec \beta^2 \les
\eps'_\sharp (\#k', \omega' ) \mu'_\sharp(\#k', \omega' ) -
\xi'^2_\sharp
(\#k', \omega' ) \ris -1 \ric \le \beta^2 - 1\ri \Delta_\sharp (\#k', \omega' ) }\\
\vspace{-8pt} \\ \displaystyle{ \eps^g_\sharp (\#k', \omega' ) = 2
\eps'_\sharp (\#k', \omega' ) \,\xi'_\sharp (\#k', \omega' )
 \beta \le 1 - \beta^2 \ri \Delta_\sharp (\#k', \omega' )} \\ \vspace{-8pt} \\
\eps^z_\sharp (\#k', \omega' ) = \eps'_\sharp (\#k', \omega' )
\end{array} \right\}, \qquad \l{e_cr}\\  && \nonumber \\
&& \left. \begin{array}{l} \displaystyle{\xi^t_\sharp (\#k', \omega'
) = \xi'_\sharp (\#k', \omega' ) \lec \beta^2 \le \eps'_\sharp
(\#k', \omega' ) \mu'_\sharp  (\#k', \omega' ) - \xi'^2_\sharp
(\#k', \omega' ) \ris + 1 \ric \le 1 - \beta^2 \ri \Delta_\sharp (\#k', \omega' )} \\
\vspace{-8pt} \\ \xi^g_\sharp (\#k', \omega' ) = \beta \Bigg(
\displaystyle{ \les \eps'_\sharp (\#k', \omega' ) \mu'_\sharp (\#k',
\omega' ) - 1 \ris \les 1 - \beta^2 \eps'_\sharp (\#k', \omega' )
\mu'_\sharp (\#k', \omega' ) \ris }
\\ \hspace{20mm}\displaystyle{ - \xi'^2_\sharp (\#k',
\omega' ) \lec \beta^2 \les \xi'^2_\sharp (\#k', \omega' ) - 2
\eps'_\sharp (\#k', \omega' ) \mu'_\sharp (\#k', \omega' ) - 1
\ris - 1 \ric }\Bigg) \Delta_\sharp (\#k', \omega' )
 \\\vspace{-8pt}  \\
\xi^z_\sharp (\#k', \omega' ) = \xi'_\sharp (\#k', \omega' )
\end{array} \right\}, \l{x_cr} \\ && \nonumber \\
&& \left. \begin{array}{l} \mu^t_\sharp (\#k', \omega' ) =
\displaystyle{ \mu'_\sharp (\#k', \omega' ) \lec \beta^2 \les
\eps'_\sharp (\#k', \omega' ) \mu'_\sharp (\#k', \omega' ) -
\xi'^2_\sharp (\#k', \omega' ) \ris
-1 \ric \le \beta^2 - 1\ri \Delta_\sharp (\#k', \omega' ) }\\ \vspace{-8pt} \\
\displaystyle{ \mu^g_\sharp (\#k', \omega' ) = 2 \mu'_\sharp (\#k',
\omega' ) \,\xi'_\sharp (\#k', \omega' )
 \beta \le 1 - \beta^2 \ri \Delta_\sharp (\#k', \omega' )} \\ \vspace{-8pt} \\
\mu^z_\sharp (\#k', \omega' ) = \mu'_\sharp (\#k', \omega' )
\end{array} \right\}, \l{m_cr}
\end{eqnarray}
with
\begin{eqnarray}
\nonumber \frac{1}{\Delta_\sharp (\#k', \omega' )} &=&    1 - 2
\beta^2 \les \eps'_\sharp (\#k', \omega' ) \mu'_\sharp (\#k',
\omega' ) +
\xi'^2_\sharp (\#k', \omega' ) \ris \\
&&+ \beta^4 \les \eps'_\sharp (\#k', \omega' ) \mu'_\sharp (\#k',
\omega' ) - \xi'^2_\sharp (\#k', \omega' ) \ris^2.
\end{eqnarray}
The constitutive relations \r{cr1} in $\Sigma$ reduce to those in
$\Sigma'$ specified by \r{cr_FD_1} in the limit $v \rightarrow 0$.
We note  the similarly of these constitutive relations \r{cr1} to
the tensor formulation that is used in plasma physics (Melrose \&
McPhedran, 1991). Interestingly, the  constitutive dyadics
\r{chi_dyadic} have the same form as that ascribed to a  Faraday
chiral medium (FCM) (Weiglhofer \& Lakhtakia 1998). Hitherto, FCMs
have been conceptualized as homogenized composite mediums
(Weiglhofer \& Lakhtakia 1998; Engheta \emph{et al.} 1992), arising
from blending together an isotropic chiral medium with either a
magnetically biased ferrite (Weiglhofer \emph{et al.} 1998) or a
magnetically biased plasma (Weiglhofer \& Mackay 2000). Through the
homogenization process, the natural optical activity of isotropic
chiral mediums (Lakhtakia 1994b) is combined with the Faraday
rotation exhibited by gyrotropic mediums (Lax \& Button 1962).

For the spatially local medium represented by the $\Sigma'$
constitutive relations \r{cr_FD_2}, the corresponding constitutive
relations in $\Sigma$ are delivered from \r{cr1} as
\begin{equation}
\left.
\begin{array}{l}
 \#D (\#r, \omega) = \epso\, \=\eps ( \omega' ) \.\#E (\#r, \omega ) +
 i \sqrt{\epso \muo} \, \=\xi ( \omega' )
  \.
 \#H(\#r, \omega)
\\ \vspace{-3mm} \\
\#B (\#r, \omega) = - i \sqrt{\epso \muo} \, \=\xi ( \omega' ) \.
\#E (\#r, \omega) + \muo\, \=\mu ( \omega' ) \.\#H (\#r, \omega)
\end{array}
\label{cr2} \right\},
\end{equation}
with the 3$\times$3 constitutive dyadics defined as in
\r{chi_dyadic}, but with no dependency on $\#k'$.

\begin{figure}[!ht]
\centering \psfull\epsfig{file=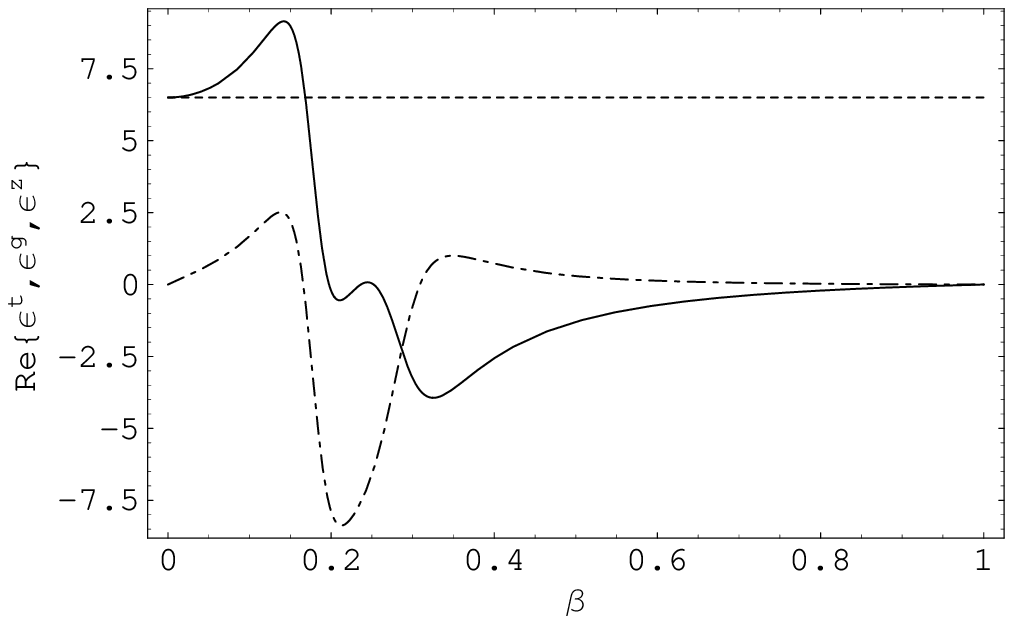,width=2.8in} \hspace{10mm}
\epsfig{file= 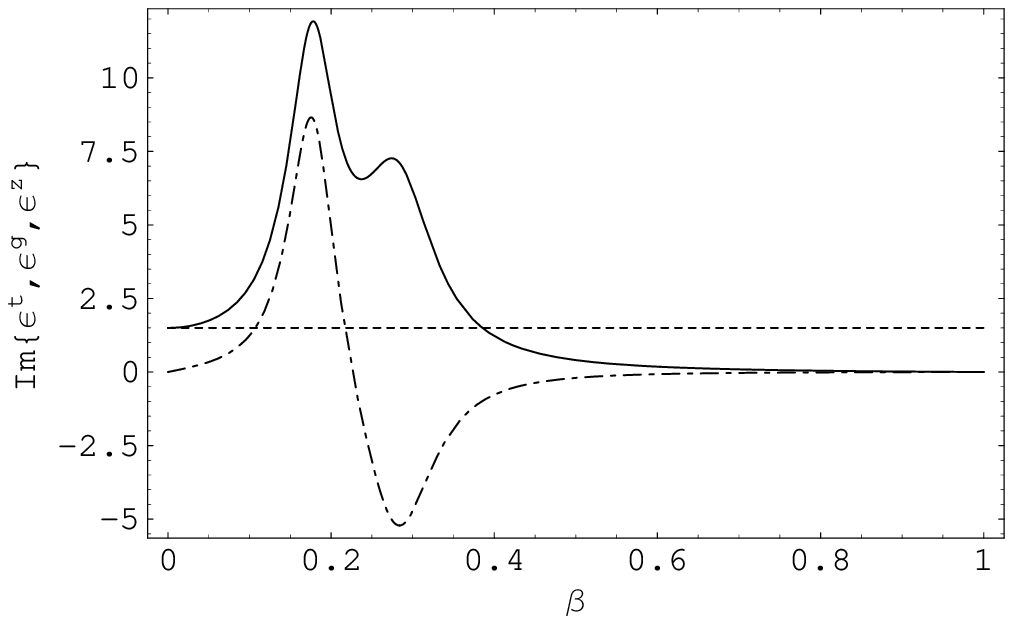,width=2.8in}\\ \vspace{10mm}
\psfull\epsfig{file=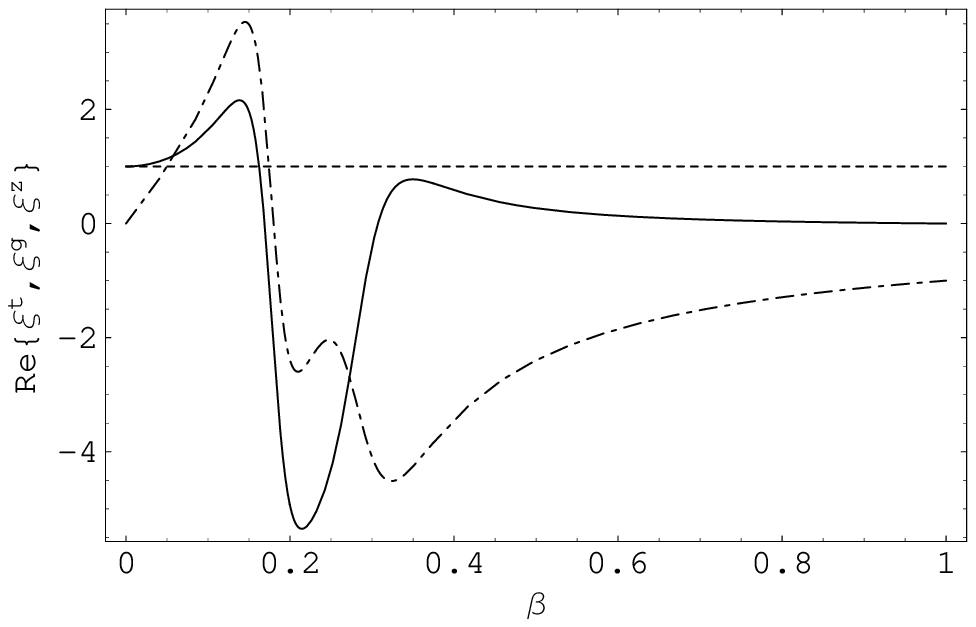,width=2.8in} \hspace{10mm}
\epsfig{file= 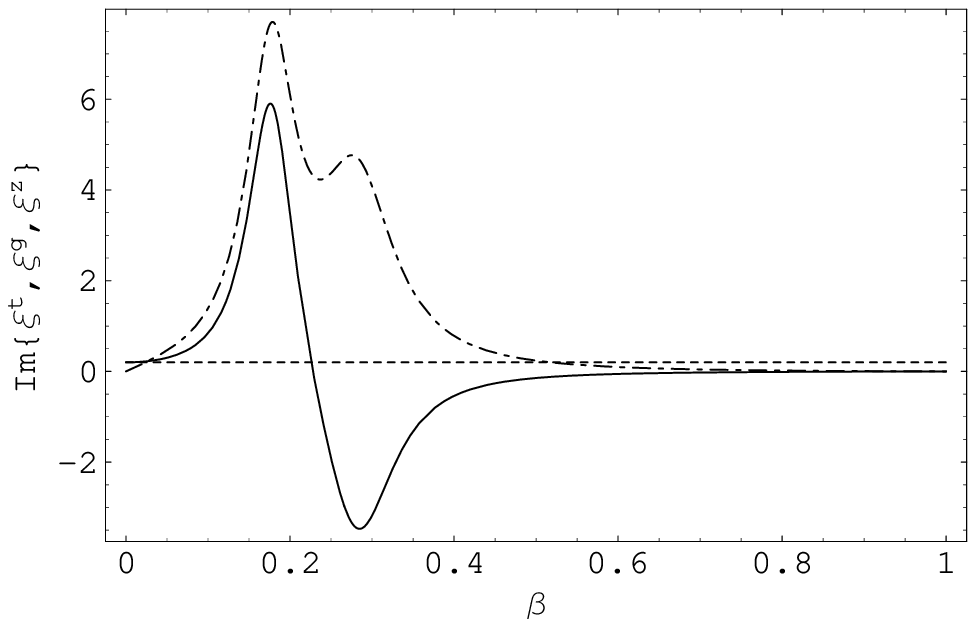,width=2.8in}
\\ \vspace{10mm}
\psfull\epsfig{file=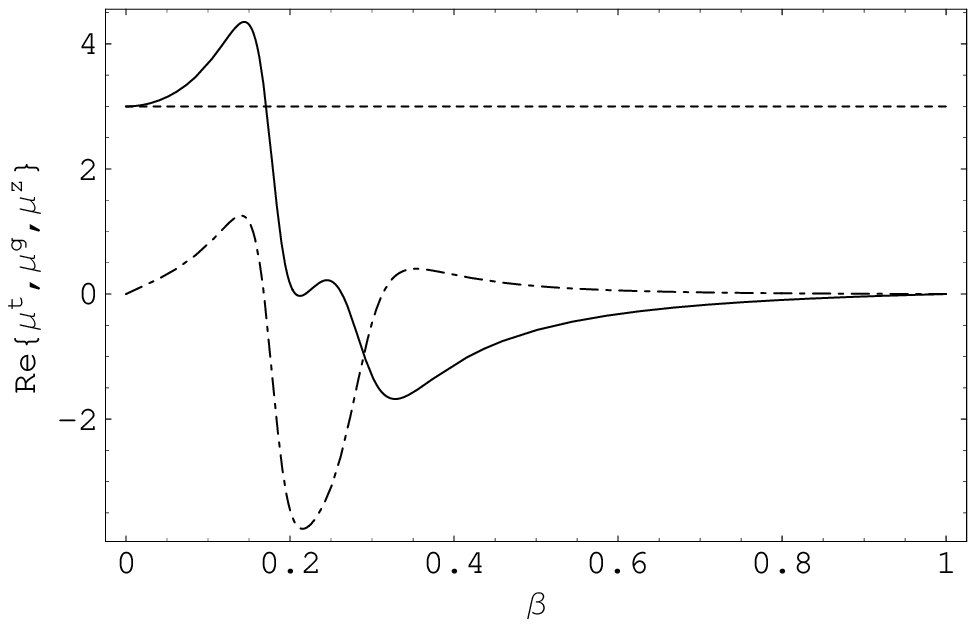,width=2.8in} \hspace{10mm}
\epsfig{file= 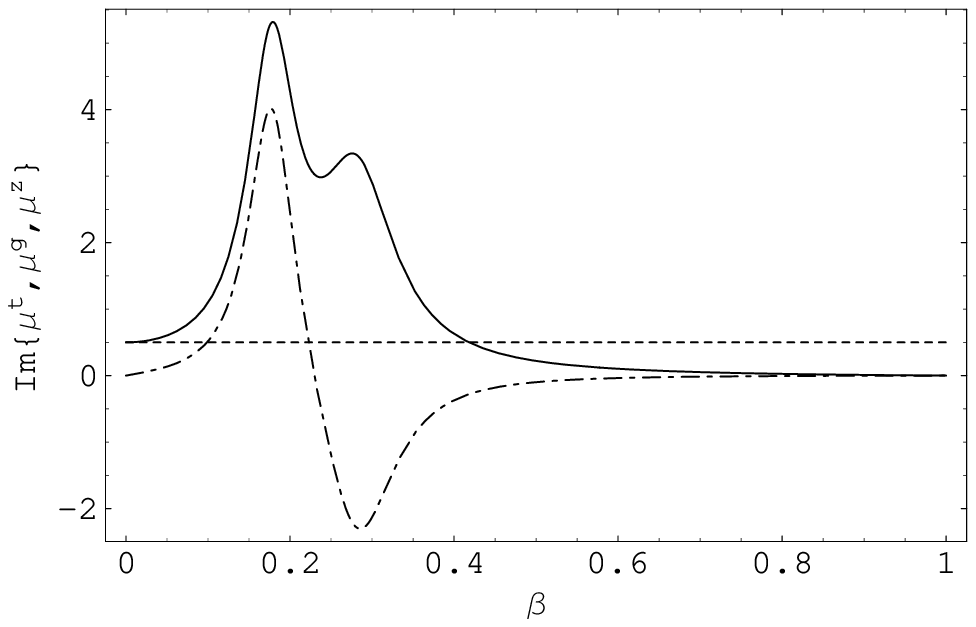,width=2.8in}
  \caption{\label{fig1} Real (left) and imaginary (right) parts of the constitutive
  parameters $\chi^t$, $\chi^g$ and $\chi^z$, where $\chi \in \lec \eps,
  \xi, \mu \ric$, plotted against the
relative speed $\beta = v/\co$. Solid curves represent $\chi$,
dashed curves represent $\chi_{{}_z}$, and
   broken dashed curves represent $\chi_{{}_g}$.
Constitutive parameters in $\Sigma'$: $\eps'_{\mbox{\tiny{}}} =
6.5 + i 1.5 $, $\xi'_{\mbox{\tiny{}}} =  1 + i 0.2 $, and
$\mu'_{\mbox{\tiny{}}} =  3.0 + i 0.5 $. }
\end{figure}

As an illustrative example, let us consider the case of the medium
specified by the  $\Sigma'$ constitutive parameters
$\eps'_{\mbox{\tiny{}}} =  6.5 + i 1.5 $, $\xi'_{\mbox{\tiny{}}} = 1
+ i 0.2 $, and $\mu'_{\mbox{\tiny{}}} = 3.0 + i 0.5 $.  The
corresponding constitutive parameters in $\Sigma$, as specified in
\r{e_cr}--\r{m_cr}, are plotted in Figure~\ref{fig1} against the
relative speed $\beta \in \les 0,1 \ri$. The parameters are
constrained such that $ \mbox{Re} \lec \chi^t, \chi^z \ric
\rightarrow \mbox{Re} \lec \chi'_{\mbox{\tiny{}}} \ric$, $ \mbox{Im}
\lec \chi^t, \chi^z \ric \rightarrow \mbox{Im} \lec
\chi'_{\mbox{\tiny{}}} \ric$ and $ \left| \chi^g \right| \rightarrow
0$ as $\beta \rightarrow 0$ for $\chi = \eps, \mu$ and $\xi$. In
Figure~\ref{fig1}, whereas $\mbox{Re} \lec \chi^t, \chi^g \ric $ and
$\mbox{Im} \lec \chi^t, \chi^g \ric $ become vanishingly small as
$\beta$ approaches unity for $\chi = \eps$ and $\mu$, as do
$\mbox{Re} \lec \xi^t \ric$ and $\mbox{Im} \lec \xi^t, \xi^g \ric$,
this is not the case for $\mbox{Re} \lec \xi^g \ric$. The parameters
$\chi^z$ are independent of $\beta$ for $\chi = \eps, \mu$ and
$\xi$.

\section{Planewave propagation} \l{pw_section}

 Let us now consider the propagation of plane waves in spatially local  mediums
 of the chosen kind.
A plane wave characterized in frame $\Sigma$ by the wavevector
$\#k$ and angular frequency $\omega$ is related to a plane wave
characterized by  the wavevector $\#k'$ and angular frequency
$\omega'$   in frame $\Sigma'$ by the relations \r{phase_invar}.
The Doppler shift and aberration arising from the transformation
from $\Sigma'$ to $\Sigma$  have been explored previously (Engheta
\emph{et al.} 1989). In the remainder of this section, we exploit
the frequency--domain constitutive relations \r{cr_FD_2} and
\r{cr2} for the two frames to consider planewave propagation and,
in particular, investigate the phenomenon of negative phase
velocity for the isotropic chiral medium moving at constant
velocity. A central element in the analysis is the introduction of
Beltrami and Beltrami--like fields.

 Planewave propagation  in reference frame $\Sigma'$, with field phasors of
 the form
\begin{equation}
\left.\begin{array}{l}
\#E'(\#r', \omega') = \#E'_{\,0} (\omega') \, \exp (i  \, \#k' \. \#r' )\\[5pt]
\#H'(\#r', \omega') = \#H'_{\,0} (\omega') \, \exp (i  \, \#k' \.
\#r' )
\end{array}\right\},
\l{pw'}
\end{equation}
  is a well--documented matter (Lakhtakia 1994b). It is both mathematically
  expedient and physically insightful to
  implement the Bohren transform and introduce the Beltrami  field
  phasors (Bohren 1974)
\begin{equation}
\l{Q12def} \left. \begin{array}{l}
 \#{\mathbb{Q}}'_{\,1} (\#r', \omega')  = \displaystyle{ \frac{1}{2} \les \#E' (\#r', \omega')
  + i \eta'(\omega')  \,
\#H' (\#r', \omega')\,
 \ris  }\vspace{4pt}\\
\#{\mathbb{Q}}'_{\,2} (\#r', \omega') = \displaystyle{ \frac{1}{2}
\les \, \#H' (\#r', \omega') + \frac{i}{\eta'(\omega') }\, \#E'
(\#r', \omega') \, \ris }\end{array} \right\},
\end{equation}
with the intrinsic impedance
\begin{equation}
\eta'(\omega')  = \les \frac{ \muo \mu'_{\mbox{\tiny{}}} ( \omega' )
}{\epso \eps'_{\mbox{\tiny{}}} (\omega')} \ris^{1/2}.
\end{equation}
 Thereby, the  frequency--domain
Maxwell curl postulates in reference frame $\Sigma'$, namely
\begin{equation}
\left.
\begin{array}{l}
\displaystyle{
\nabla' \times \#E' (\#r', \omega') - i \omega'   \#B' (\#r', \omega') = \#0} \\
\vspace{-3mm}\\
\displaystyle{\nabla' \times \#H'(\#r', \omega') + i \omega'
\#D'(\#r', \omega') = \#0}
\end{array}
\right\}, \l{Sigma_p_FD}
\end{equation}
may be recast as two uncoupled  first--order differential
equations, which yield
\begin{equation}
i \#k' \times \#{\mathbb{Q}}'_{\,\ell}\,(\#r',\omega') + (-1)^\ell
\ko' \tilde{k}'_\ell \,\#{\mathbb{Q}}'_{\,\ell}\,(\#r',\omega') =
\#0\, , \hspace{20mm} \le \ell = 1,2 \ri, \l{B12}
\end{equation}
for plane waves \r{pw'}. Regardless of the direction of propagation, two
wavevectors $\#k' \in \lec \#k'_\ell \ric$ with $ \#k'_\ell =
k'_\ell \hat{\#k'_\ell}$ and corresponding wavenumbers $k'_\ell =
\ko' \tilde{k}'_\ell$ are supported, where (Lakhtakia 1994b)
\begin{equation}
\left.
\begin{array}{l}
\tilde{k}'_1 = \sqrt{\eps'_{\mbox{\tiny{}}}  (\omega')\,
\mu'_{\mbox{\tiny{}}} (\omega') } - \xi'_{\mbox{\tiny{}}}
(\omega')\\ \vspace{-8pt} \\
\tilde{k}'_2 = \sqrt{\eps'_{\mbox{\tiny{}}}  (\omega')\,
\mu'_{\mbox{\tiny{}}} (\omega')} + \xi'_{\mbox{\tiny{}}} (\omega')
\end{array} \right\}
\end{equation}
and $\ko' = \omega' \sqrt{\epso \muo}$.

The nonreciprocal bianisotropic nature (Krowne 1984) of the medium
specified by \r{cr2} in reference frame $\Sigma$ leads to more
complicated planewave characteristics than in $\Sigma'$. Following
the strategy used for frame $\Sigma'$, it is helpful to utilize
the field phasors
\begin{equation}
\l{Q12def_x} \left. \begin{array}{l}
 \#{\mathbb{Q}}_{\,1} (\#r, \omega)  = \displaystyle{ \frac{1}{2} \les \#E (\#r, \omega)
  + i \eta' (\omega') \,
\#H (\#r, \omega)\,
 \ris  }\vspace{4pt}\\
\#{\mathbb{Q}}_{\,2} (\#r, \omega) = \displaystyle{ \frac{1}{2} \les
\, \#H (\#r, \omega) + \frac{i}{\eta' (\omega')} \, \#E (\#r,
\omega) \, \ric }\end{array} \right\}.
\end{equation}
 This enables the  frequency--domain Maxwell curl postulates in frame $\Sigma$, namely
\begin{equation}
\left.
\begin{array}{l}
\displaystyle{
\nabla \times \#E (\#r, \omega) - i \omega   \#B (\#r, \omega) = \#0} \\
\vspace{-3mm}\\
\displaystyle{\nabla \times \#H(\#r, \omega) + i \omega \#D(\#r,
\omega) = \#0}
\end{array}
\right\}, \l{Sigma_FD}
\end{equation}
to be  decoupled as
\begin{equation}
i  \#k \times \#{\mathbb{Q}}_{\,\ell}\,(\#r,\omega) + (-1)^\ell
\,\ko \, \tilde{\=\kappa}_\ell (\omega')
\.\#{\mathbb{Q}}_{\,\ell}\,(\#r,\omega) = \#0 , \hspace{20mm} \le
\ell = 1,2 \ri, \l{BB12}
\end{equation}
for plane waves
\begin{equation}
\left.\begin{array}{l}
\#E(\#r) = \#E_{\,0}\, \exp (i  \, \#k \. \#r )\\[5pt]
\#H(\#r) = \#H_{\,0} \, \exp (i  \, \#k \. \#r )
\end{array}\right\},
\l{pw}
\end{equation}
with wavevector $\#k = k \hat{\#k}$ and  $\ko = \omega \sqrt{\epso
\muo}$.
  The 3$\times$3  dyadics in \r{BB12} are given as
\begin{equation}
  \tilde{\=\kappa}_{\,\ell} (\omega') = \tilde{\kappa}^t_\ell ( \omega' ) \, \=I - i
\tilde{\kappa}^{g}_\ell ( \omega' )\, \hat{\#v} \times \=I + \les
\tilde{\kappa}^{z}_\ell ( \omega' ) - \tilde{\kappa}^t_\ell (
\omega' ) \ris \hat{\#v} \, \hat{\#v}\,,  \hspace{8mm} (\ell= 1,2
),\l{kappa_dyadic}
\end{equation}
with
\begin{equation}
\left. \begin{array}{l} \tilde{\kappa}^\nu_1 =   \displaystyle{
\les\frac{ \mu' (\omega')}{\eps' (\omega')}\ris^{1/2}  \eps^\nu
(\omega') - \xi^\nu (\omega') } \equiv  \displaystyle{\les
\frac{\eps' (\omega') }{\mu' (\omega')}\ris^{1/2}
 \mu^\nu (\omega') - \xi^\nu (\omega')  }\vspace{4mm} \\
\tilde{\kappa}^\nu_2 =  \displaystyle{\les \frac{ \mu'
(\omega')}{\eps' (\omega')}\ris^{1/2} \eps^\nu (\omega') + \xi^\nu
(\omega') }\equiv  \displaystyle{ \les \frac{\eps' (\omega')
}{\mu' (\omega')} \ris^{1/2} \mu^\nu (\omega') + \xi^\nu (\omega')
}
\end{array} \l{kappa_def}
\right\}, \hspace{5mm} (\nu = t, g, z).
\end{equation}
  The unprimed constitutive parameters in \r{kappa_def} are
defined as in \r{e_cr}--\r{m_cr}, but with no dependency on $\#k'$.
Whereas $\#{\mathbb{Q}}'_{\,\ell}\,(\#r',\omega')$ are Beltrami
field phasors (Lakhtakia 1994a),
$\#{\mathbb{Q}}_{\,\ell}\,(\#r,\omega)$ should be called
Beltrami--like field phasors.

The dispersion relations
\begin{equation}
\mbox{det} \les i \#k \times \=I + (-1)^\ell \ko \,
\tilde{\=\kappa}_\ell (\omega') \ris = 0, \hspace{25mm} \le \ell =
1,2 \ri, \l{disp}
\end{equation}
arise immediately from \r{BB12}. For an arbitrary direction of
propagation specified by the  relative orientation angle $\theta =
\cos^{-1} \hat{\#v} \. \hat{\#k}$, the dispersion relations \r{disp}
may be expressed as the pair of quadratic equations
\begin{equation}
 a_\ell \,\tilde{k}^2 + b_\ell\, \tilde{k} + c_\ell =0\,, \hspace{25mm} \le \ell = 1,2 \ri, \l{quad}
\end{equation}
wherein the relative wavenumber $\tilde{k} = k / \ko$. The
coefficients in these equations are given as
\begin{equation}
\left. \begin{array}{l} a_1 =  1+ \beta^2 \Bigg( \lec \les
\sqrt{\eps'_{\mbox{\tiny{}}}(\omega') \, \mu'_{\mbox{\tiny{}}}
(\omega') } + \xi'_{\mbox{\tiny{}}} (\omega') \ris^2 \cos^2 \theta
+ \sin^2 \theta \ric \\  \hspace{10mm} \times \lec \beta^2 \les
\sqrt{\eps'_{\mbox{\tiny{}}}(\omega') \, \mu'_{\mbox{\tiny{}}}
(\omega') } - \xi'_{\mbox{\tiny{}}} (\omega') \ris^2 -1 \ric   -
\les \sqrt{\eps'_{\mbox{\tiny{}}}(\omega') \,
\mu'_{\mbox{\tiny{}}} (\omega') } -
\xi'_{\mbox{\tiny{}}} (\omega')  \ris^2 \Bigg)  \\
b_1 = 2 \beta \cos \theta  \lec 1- \les
\sqrt{\eps'_{\mbox{\tiny{}}}(\omega') \, \mu'_{\mbox{\tiny{}}}
(\omega') } + \xi'_{\mbox{\tiny{}}} (\omega')  \ris^2 \ric \\
\hspace{10mm} \times \lec \beta^2 \les
\sqrt{\eps'_{\mbox{\tiny{}}}(\omega')
\, \mu'_{\mbox{\tiny{}}} (\omega') } - \xi'_{\mbox{\tiny{}}} (\omega')  \ris^2 -1 \ric\\
c_1 =  \beta^2 \lec \les \eps'_{\mbox{\tiny{}}}(\omega') \,
\mu'_{\mbox{\tiny{}}} (\omega') - \xi'^2_{\mbox{\tiny{}}}
(\omega') \ris^2 +1 \ric - \les
\sqrt{\eps'_{\mbox{\tiny{}}}(\omega') \, \mu'_{\mbox{\tiny{}}}
(\omega') } - \xi'_{\mbox{\tiny{}}} (\omega') \ris^2 \le \beta^4 +
1 \ri
\end{array}
\right\}
\end{equation}
and
\begin{equation}
\left. \begin{array}{l} a_2 = - 1 +  \beta^2 \Bigg( \lec \les
\sqrt{\eps'_{\mbox{\tiny{}}}(\omega') \, \mu'_{\mbox{\tiny{}}}
(\omega') } - \xi'_{\mbox{\tiny{}}} (\omega')  \ris^2 \cos^2
\theta + \sin^2 \theta \ric \\ \hspace{10mm} \times \lec 1-\beta^2
\les \sqrt{\eps'_{\mbox{\tiny{}}}(\omega') \,
\mu'_{\mbox{\tiny{}}} (\omega') } + \xi'_{\mbox{\tiny{}}}
(\omega')  \ris^2
 \ric  + \les \sqrt{\eps'_{\mbox{\tiny{}}}(\omega') \, \mu'_{\mbox{\tiny{}}}
(\omega') } +
\xi'_{\mbox{\tiny{}}} (\omega')  \ris^2 \Bigg)  \\
b_2 = 2 \beta \cos \theta \Bigg[  \les
\sqrt{\eps'_{\mbox{\tiny{}}}(\omega') \, \mu'_{\mbox{\tiny{}}}
(\omega') } + \xi'_{\mbox{\tiny{}}} (\omega')  \ris^2 \\
\hspace{10mm} \times \Bigg( \beta^2 \lec
   \les \sqrt{\eps'_{\mbox{\tiny{}}}(\omega') \,
\mu'_{\mbox{\tiny{}}} (\omega') } - \xi'_{\mbox{\tiny{}}}
(\omega')  \ris^2  - 1 \ric -1 \Bigg) + 1 \Bigg]
\\
c_2 = \les \sqrt{\eps'_{\mbox{\tiny{}}}(\omega') \,
\mu'_{\mbox{\tiny{}}} (\omega') } + \xi'_{\mbox{\tiny{}}}
(\omega') \ris^2 \le \beta^4 + 1 \ri  -  \beta^2 \lec \les
\eps'_{\mbox{\tiny{}}}(\omega') \, \mu'_{\mbox{\tiny{}}} (\omega')
- \xi'^2_{\mbox{\tiny{}}} (\omega') \ris^2 + 1 \ric
\end{array}
\right\}.
\end{equation}
Thus, four relative  wavenumbers $\tilde{k} \in \lec \tilde{k}_1,
\tilde{k}_2, \tilde{k}_3, \tilde{k}_4 \ric$ emerge as the roots of
\r{quad}. While these may be straightforwardly extracted from
\r{quad},
 explicit algebraic
representations of the wavenumbers are generally cumbersome. The
following two special cases are noteworthy exceptions. For
propagation parallel to the direction of translation (i.e.,
$\hat{\#k} = \hat{\#v}$) we have
\begin{equation}
\left. \begin{array}{l}
 \tilde{k}_1 = \displaystyle{  \frac{  \sqrt{\eps'_{\mbox{\tiny{}}}
(\omega') \mu'_{\mbox{\tiny{}}} (\omega')} \le 1-\beta^2 \ri +
\beta \les 1 - \eps'_{\mbox{\tiny{}}} (\omega')
\mu'_{\mbox{\tiny{}}} (\omega') + \xi'^2_{\mbox{\tiny{}}}
(\omega') \ris - \xi'_{\mbox{\tiny{}}} (\omega') \le 1+\beta^2
\ri}{1-\beta \lec \beta \les \eps'_{\mbox{\tiny{}}} (\omega')
\mu'_{\mbox{\tiny{}}} (\omega')
 - \xi'^2_{\mbox{\tiny{}}} (\omega') \ris - 2 \xi'_{\mbox{\tiny{}}} (\omega') \ric }}\\
\vspace{-6pt} \\ \tilde{k}_2 = \displaystyle{
\frac{\sqrt{\eps'_{\mbox{\tiny{}}} (\omega') \mu'_{\mbox{\tiny{}}}
(\omega')} \le 1-\beta^2 \ri +\beta \les 1 -
\eps'_{\mbox{\tiny{}}} (\omega') \mu'_{\mbox{\tiny{}}}(\omega') +
\xi'^2_{\mbox{\tiny{}}} (\omega') \ris +
\xi'_{\mbox{\tiny{}}}(\omega') \le 1+\beta^2 \ri }{1-\beta \lec
\beta \les \eps'_{\mbox{\tiny{}}}(\omega')
\mu'_{\mbox{\tiny{}}}(\omega') - \xi'^2_{\mbox{\tiny{}}}(\omega')
\ris - 2 \xi'_{\mbox{\tiny{}}} (\omega') \ric }}
\\ \vspace{-6pt} \\
\tilde{k}_3 = \displaystyle{  \frac{ -
\sqrt{\eps'_{\mbox{\tiny{}}}(\omega') \mu'_{\mbox{\tiny{}}}
(\omega')} \le 1- \beta^2 \ri +\beta \les 1 -
\eps'_{\mbox{\tiny{}}} (\omega') \mu'_{\mbox{\tiny{}}} (\omega') +
\xi'^2_{\mbox{\tiny{}}} (\omega') \ris + \xi'_{\mbox{\tiny{}}}
(\omega') \le 1+\beta^2 \ri}{1 - \beta \lec \beta \les
\eps'_{\mbox{\tiny{}}} (\omega') \mu'_{\mbox{\tiny{}}} (\omega') -
\xi'^2_{\mbox{\tiny{}}} (\omega') \ris - 2 \xi'_{\mbox{\tiny{}}}(\omega') \ric }}\\
\vspace{-6pt} \\
\tilde{k}_4 = \displaystyle{ \frac{ -
\sqrt{\eps'_{\mbox{\tiny{}}}(\omega') \mu'_{\mbox{\tiny{}}}
(\omega')} \le 1-\beta^2 \ri + \beta \les 1 -
\eps'_{\mbox{\tiny{}}}(\omega') \mu'_{\mbox{\tiny{}}}(\omega') +
\xi'^2_{\mbox{\tiny{}}}(\omega') \ris -
\xi'_{\mbox{\tiny{}}}(\omega') \le 1+ \beta^2 \ri}{1-\beta \lec
\beta \les \eps'_{\mbox{\tiny{}}}(\omega')
\mu'_{\mbox{\tiny{}}}(\omega') - \xi'^2_{\mbox{\tiny{}}}(\omega')
\ris - 2 \xi'_{\mbox{\tiny{}}} (\omega') \ric }}
\end{array}
\right\},
\end{equation}
whereas  the relative wavenumbers are
delivered as
\begin{equation}
\left. \begin{array}{l} \tilde{k}_1 = \displaystyle{\lec \frac{
\les \sqrt{\eps'_{\mbox{\tiny{}}} (\omega') \mu'_{\mbox{\tiny{}}}
(\omega')} - \xi'_{\mbox{\tiny{}}} (\omega') \ris^2 - \beta^2 }{1-
\beta^2} \ric^{1/2}}\\ \vspace{-6pt} \\
\tilde{k}_2 = \displaystyle{\lec \frac{ \les
\sqrt{\eps'_{\mbox{\tiny{}}} (\omega') \mu'_{\mbox{\tiny{}}}
(\omega')}
+ \xi'_{\mbox{\tiny{}}} (\omega') \ris^2  - \beta^2 }{1- \beta^2} \ric^{1/2}}\\
\vspace{-6pt} \\ \tilde{k}_3 = - \tilde{k}_1
\\\tilde{k}_4 = - \tilde{k}_2
\end{array}
\right\}
\end{equation}
for propagation perpendicular to the direction of translation
(i.e., $\hat{\#k} \. \hat{\#v} = 0 $).

By way of numerical illustration, let us return to the constitutive parameters
used for Figure~\ref{fig1}. The corresponding relative
wavenumbers in $\Sigma$, computed as the roots of \r{quad}, are
plotted in Figure~\ref{fig2} against relative speed $\beta \in
\les 0,1 \ri$ and wavevector orientation angle  $\theta \in \les
0, \pi \ri$. For clarity, the wavenumbers in Figure~\ref{fig2} are
ordered such that $\mbox{Im} \lec \tilde{k}_1 \ric$ $>$ $\mbox{Im}
\lec \tilde{k}_2 \ric$ $>$ $\mbox{Im} \lec \tilde{k}_3 \ric$ $>$
$\mbox{Im} \lec \tilde{k}_4 \ric$. Notice that $\mbox{Im} \lec
\tilde{k}_{1,2} \ric$ $> 0$, whereas $\mbox{Im} \lec
\tilde{k}_{3,4} \ric$ $ < 0$. At $\beta = 0$, the wavenumbers are
independent of $\theta$. As $\beta$ increases from zero, the
dependencies of the wavenumbers upon $\theta$ are observed to be
highly asymmetric with respect to $\theta = \pi/2$. In the limit
$\beta \rightarrow 1 $, the $\theta$--dependencies of the real
parts of the wavenumbers
 become antisymmetric  relative to $\theta = \pi/2$, whereas the
 $\theta$--dependencies  of the imaginary parts of
the wavenumbers become
  symmetric relative to $\theta = \pi/2$.

In relation to planewave propagation,  a topic of considerable
current interest is whether the phase velocity is negative or
positive (Lakhtakia \emph{et al.} 2003). Negative phase velocity
(NPV) is closely related to the phenomenon of negative refraction
(Ramakrishna 2005). Planewave propagation with NPV in $\Sigma$ is
signified by (Mackay \& Lakhtakia 2004b)
\begin{equation}
\l{NPV-def}
{\rm Re}\lec\#k\ric \. \#P (\#r, \omega) < 0,
\end{equation}
where $\#P (\#r, \omega) $ is the time--averaged Poynting vector;
conversely, positive phase velocity (PPV) in $\Sigma$ is signified
by
\begin{equation}
{\rm Re}\lec\#k\ric \. \#P (\#r, \omega) > 0.
\end{equation}
In $\Sigma'$, NPV is signified by ${\rm Re}\lec \#k'\ric \.\#P'
(\#r', \omega') < 0$ and PPV by
 ${\rm Re}\lec \#k'\ric \.\#P' (\#r', \omega')> 0$.
Issues concerning  NPV propagation for isotropic chiral mediums
(Mackay 2005) and FCMs arising as homogenized composite mediums
(Mackay \& Lakhtakia 2004b) have been reported previously.

For the medium of interest here, NPV propagation occurs in
$\Sigma'$ provided that (Mackay 2005)
\begin{equation}
\left.
\begin{array}{l}
\mbox{Re} \lec \sqrt{\eps'_{\mbox{\tiny{}}} (\omega')\,
\mu'_{\mbox{\tiny{}}} (\omega') }
 - \xi'_{\mbox{\tiny{}}} (\omega') \ric \times \mbox{Re}
\lec \sqrt{\frac{ \displaystyle \eps'^*_{\mbox{\tiny{}}}
(\omega')}{ \displaystyle \mu'^*_{\mbox{\tiny{}}} (\omega')}} \,
\ric < 0
  \qquad
\mbox{for} \qquad k' = \ko' \tilde{k}'_{1,3} \\
\\
\mbox{Re} \lec \sqrt{\eps'_{\mbox{\tiny{}}} (\omega')\,
\mu'_{\mbox{\tiny{}}} (\omega') } + \xi'_{\mbox{\tiny{}}}
(\omega') \ric \times \mbox{Re} \lec \sqrt{\frac{ \displaystyle
\eps'^*_{\mbox{\tiny{}}} (\omega')}{ \displaystyle
\mu'^*_{\mbox{\tiny{}}} (\omega')}} \, \ric < 0
  \qquad
\mbox{for} \qquad k' = \ko' \tilde{k}'_{2,4}
\end{array}
\right\}. \l{NPV_cond_ICM}
\end{equation}
For the same medium, by virtue of \r{NPV-def}, NPV propagation occurs in $\Sigma$ provided
that
\begin{equation}
\Omega(\tilde{k} ) < 0\,, \l{w_NPV}
\end{equation}
the form of the  real--valued NPV parameter $\Omega (\tilde{k})$
being provided in Appendix~1.

Let us return to the numerical example considered in
Figures~\ref{fig1} and \ref{fig2}. The $\beta\theta$--regimes of
NPV and PPV, as determined by evaluating  $\Omega (\tilde{k})$ for
$\tilde{k} \in \lec \tilde{k}_1, \tilde{k}_2, \tilde{k}_3,
\tilde{k}_4 \ric$, are mapped in Figure~\ref{fig3} with respect to
the relative speed $\beta \in \les 0,1 \ri$ and wavevector
orientation angle $\theta \in \les 0, \pi \ri$. The medium clearly
does not support NPV propagation when $\beta = 0$; i.e., all plane
waves in $\Sigma'$ must be of the PPV kind. As $\beta$ increases,
the $\beta\theta$--regimes supporting NPV propagation in $\Sigma$
emerge  in the range $\pi/2 < \theta < \pi$ for the  relative
wavenumbers $\tilde{k}_{1,2}$, and in the range
 $0 < \theta < \pi/2$ for the
relative wavenumbers $\tilde{k}_{3,4}$.

Finally in this section, we note that an alternative derivation of
the  dispersion relations \r{disp} in $\Sigma$ may be developed via the
Lorentz transformation of the corresponding dispersion
relations in $\Sigma'$. NPV arises when this Lorentz transformation brings about
a change of sign in the angular frequency.

\section{Dyadic Green functions} \l{DGF_section}

The problem of finding the (frequency--domain) field phasors
generated by a specified distribution of sources  within a linear
medium is conveniently tackled by means of dyadic Green functions
(DGFs) (Tai 1994). For the isotropic chiral medium  in reference
frame $\Sigma'$, the DGFs are well--known (Lakhtakia 1994b).
However,  explicit
representations of DGFs are generally unavailable
for anisotropic and bianisotropic mediums (Mackay \&
Lakhtakia 2006).
 In this section we exploit the
constitutive relations derived in \S\ref{CR_section}, together
with the Beltrami--like fields introduced in \S\ref{pw_section},
to establish a convenient spectral representation of the DGFs for
the FCM described by \r{cr2}.

Let a source electric current density phasor $\#J_{\,e} (\#r,
\omega)$ and a source magnetic current density phasor $\#J_{\,m}
(\#r, \omega)$ exist, from the perspective of the non--co--moving
observer. Extending the approach adopted in \S\ref{pw_section}
wherein Beltrami--like fields are introduced to aid the planewave
analysis in $\Sigma$, we recast $\#J_{\,e,m} (\#r, \omega)$ as the
Beltrami current density phasors
\begin{equation}
 \l{W_Beltrami} \left. \begin{array}{l}
 \#{\mathbb{W}}_{\,1} (\#r, \omega)  = \displaystyle{ \frac{1}{2} \, \les i
\eta' (\omega')\, \#J_{\,e} (\#r, \omega ) - \#J_{\,m}( \#r,
\omega)\,
 \ris }\vspace{2pt} \\
 \#{\mathbb{W}}_{\,2} (\#r, \omega)= \displaystyle{\frac{1}{2} \les
   \#J_{\,e}( \#r',\omega) -  \displaystyle{\frac{i}{\eta' (\omega')}}\,
\#J_{\,m} (\#r, \omega )\,
 \ris  }\end{array} \right\}.
\end{equation}
The Beltrami--like field phasors generated by the  source terms
\r{W_Beltrami} may then be expressed in terms of the DGFs
$\={\mathbb{G}}_{\,\ell} (\#r - \#s, \omega')$ as
\begin{equation} \l{Beltrami_soln}
  \#{\mathbb{Q}}_{\,\ell} (\#r, \omega) =
\#{\mathring{\mathbb{Q}}}_{\,\ell} (\#r, \omega) + \, \int_{V} \,
\={\mathbb{G}}_{\,\ell} (\#r - \#s, \omega') \.
\#{\mathbb{W}}_{\,\ell} (\#r, \omega) \, \, d^3\#s\,, \qquad
\qquad \le \ell = 1,2 \ri,
\end{equation}
where  $V$ is the region containing the source current density
phasors. The complementary functions
$\#{\mathring{\mathbb{Q}}}_{\,1,2} (\#r, \omega)$
 are given by
\begin{equation}
\l{Q12def_xx} \left. \begin{array}{l}
 \#{\mathring{\mathbb{Q}}}_{\,1} (\#r, \omega)  = \displaystyle{ \frac{1}{2} \les \#{\mathring{E}} (\#r, \omega)
  + i \eta' (\omega') \,
\#{\mathring{H}} (\#r, \omega)\,
 \ris  }\vspace{4pt}\\
\#{\mathring{\mathbb{Q}}}_{\,2} (\#r, \omega) = \displaystyle{
\frac{1}{2} \les \, \#{\mathring{H}} (\#r, \omega) +
\frac{i}{\eta' (\omega')} \, \#{\mathring{E}} (\#r, \omega) \,
\ric }\end{array} \right\},
\end{equation}
wherein $\#{\mathring{E}} (\#r, \omega)$ and $\#{\mathring{H}}
(\#r, \omega)$ satisfy the relations
\begin{equation}
\left.
\begin{array}{l}
\displaystyle{ \nabla \times \#{\mathring{E}}(\#r, \omega) - i
\omega   \#{\mathring{B}}(\#r,
\omega) \equiv \#0} \\
\vspace{-3mm}\\
\displaystyle{\nabla \times \#{\mathring{H}}(\#r, \omega) + i
\omega \#{\mathring{D}}(\#r, \omega) \equiv \#0}
\end{array}
\right\},
\end{equation}
along with
\begin{equation}
\left.
\begin{array}{l}
\#{\mathring{D}} (\#r, \omega) = \epso\, \=\eps ( \omega' )
\.\#{\mathring{E}} (\#r, \omega ) +
 i \sqrt{\epso \muo} \, \=\xi ( \omega' )
  \.
\#{\mathring{H}}(\#r, \omega)
\\ \vspace{-3mm} \\
\#{\mathring{B}} (\#r, \omega) = - i \sqrt{\epso \muo} \, \=\xi (
\omega' ) \. \#{\mathring{E}} (\#r, \omega) + \muo\, \=\mu (
\omega' ) \.\#{\mathring{H}} (\#r, \omega)
\end{array}
\right\}.
\end{equation}
 The DGFs in \r{Beltrami_soln} are provided as the
solutions of the differential equations
\begin{equation}
\nabla \times \={\mathbb{G}}_{\,\ell}\,(\#r-\#s,\omega') +
(-1)^\ell  \,\ko\, \tilde{\=\kappa}_\ell (\omega')
\.\={\mathbb{G}}_{\,\ell}\,(\#r-\#s,\omega') = \delta (\#r -
\#s)\, \=I , \hspace{10mm} \le \ell = 1,2 \ri, \l{DGF12}
\end{equation}
with $\delta(\.)$ being the Dirac delta function. By implementing
 the spatial Fourier transforms
\begin{equation} \l{DGF_a}
\={\mathbb{G}}^\sharp_{\,\ell}\,(\#q,\omega') = \int_{\#r}
\={\mathbb{G}}_{\,\ell}\,(\#r,\omega') \, \exp \le - i\, \#q\.\#r
\ri \; d\#r\,, \hspace{10mm} \le \ell = 1,2 \ri
\end{equation}
with \r{DGF12}, the components of the  spectral DGFs
$\={\mathbb{G}}^\sharp_{\,1,2}\,(\#q,\omega')$ with respect to the
Cartesian basis vectors $\lec \hat{\#x}, \hat{\#y}, \hat{\#z}
\ric$ emerge as
\begin{equation}
\left.
\begin{array}{l}
 \les \={\mathbb{G}}^\sharp_{\,1}\,(\#q,\omega') \ris_{11}
= \les \le \#q \. \hat{\#x} \ri^2- \tilde{\kappa}^t_1
\tilde{\kappa}^z_1 \ris \Lambda_1 \vspace{2mm} \\
\les \={\mathbb{G}}^\sharp_{\,1}\,(\#q,\omega') \ris_{12} = \les
\le \#q \. \hat{\#x} \ri \le \#q \. \hat{\#y} \ri
 + i \tilde{\kappa}^z_1 \le  \tilde{\kappa}^g_1 + \#q\.\hat{\#z} \ri \ris \Lambda_1 \vspace{2mm}
\\
\les \={\mathbb{G}}^\sharp_{\,1}\,(\#q,\omega') \ris_{13} = \les
 \#q \. \hat{\#x} \le  \#q \. \hat{\#z} + \tilde{\kappa}^g_1 \ri -
  i \tilde{\kappa}^t_1  \#q \. \hat{\#y}  \ris \Lambda_1 \vspace{2mm}
  \\
\les \={\mathbb{G}}^\sharp_{\,1}\,(\#q,\omega') \ris_{21} = \les
\le \#q \. \hat{\#x} \ri \le \#q \. \hat{\#y} \ri
 - i \tilde{\kappa}^z_1 \le  \tilde{\kappa}^g_1 + \#q\.\hat{\#z} \ri \ris \Lambda_1 \vspace{2mm}
\\
\les \={\mathbb{G}}^\sharp_{\,1}\,(\#q,\omega') \ris_{22} = \les
\le \#q \. \hat{\#y} \ri^2- \tilde{\kappa}^t_1 \tilde{\kappa}^z_1
\ris \Lambda_1 \vspace{2mm}
\\
 \les \={\mathbb{G}}^\sharp_{\,1}\,(\#q,\omega')
\ris_{23} = \les  \#q \. \hat{\#y} \le \#q \. \hat{\#z}  +
\tilde{\kappa}^g_1 \ri + i  \tilde{\kappa}^t_1 \#q \. \hat{\#x}
\ris \Lambda_1 \vspace{2mm}
\\
\les \={\mathbb{G}}^\sharp_{\,1}\,(\#q,\omega') \ris_{31} = \les
 \#q \. \hat{\#x} \le  \#q \. \hat{\#z} + \tilde{\kappa}^g_1 \ri +
  i \tilde{\kappa}^t_1  \#q \. \hat{\#y}  \ris \Lambda_1 \vspace{2mm}
  \\
  \les \={\mathbb{G}}^\sharp_{\,1}\,(\#q,\omega')
\ris_{32} = \les  \#q \. \hat{\#y} \le \#q \. \hat{\#z}  +
\tilde{\kappa}^g_1 \ri - i  \tilde{\kappa}^t_1 \#q \. \hat{\#x}
\ris \Lambda_1 \vspace{2mm}
\\
\les \={\mathbb{G}}^\sharp_{\,1}\,(\#q,\omega') \ris_{33} = \les
\le \tilde{\kappa}^t_1 + \tilde{\kappa}^g_1 + \#q \. \hat{\#z} \ri
\le \tilde{\kappa}^t_1 - \tilde{\kappa}^g_1 - \#q \. \hat{\#z} \ri
\ris \Lambda_1
\end{array}
\right\},
\end{equation}
with
\begin{eqnarray}
\frac{1}{\Lambda_1} &=& \frac{1}{2} \lec \les \le \#q\.\hat{\#z}
\ri^2 - \le \#q\.\hat{\#x} \ri^2 - \le \#q\.\hat{\#y} \ri^2 \ris
\le \tilde{\kappa}^z_1 - \tilde{\kappa}^t_1 \ri + \#q\.\#q \le
\tilde{\kappa}^t_1 + \tilde{\kappa}^z_1 \ri \ric \nonumber \\ && +
\le \tilde{\kappa}^g_1 \ri^2 + 2 \tilde{\kappa}^g_1
\tilde{\kappa}^z_1 \#q\. \hat{\#z} - \le \tilde{\kappa}^t_1
\ri^2\,,
\end{eqnarray}
 and
\begin{equation}
\left. \begin{array}{l} \les
\={\mathbb{G}}^\sharp_{\,2}\,(\#q,\omega') \ris_{11} = \les - \le
\#q \. \hat{\#x} \ri^2+ \tilde{\kappa}^t_1
\tilde{\kappa}^z_2 \ris \Lambda_2 \vspace{2mm} \\
\les \={\mathbb{G}}^\sharp_{\,2}\,(\#q,\omega') \ris_{12} = \les -
\le \#q \. \hat{\#x} \ri \le \#q \. \hat{\#y} \ri
 + i \tilde{\kappa}^z_2 \le  - \tilde{\kappa}^g_2 + \#q\.\hat{\#z} \ri \ris \Lambda_2 \vspace{2mm}
\\
\les \={\mathbb{G}}^\sharp_{\,2}\,(\#q,\omega') \ris_{13} = \les
 \#q \. \hat{\#x} \le  \#q \. \hat{\#z} - \tilde{\kappa}^g_2 \ri +
  i \tilde{\kappa}^t_2  \#q \. \hat{\#y}  \ris \Lambda_2 \vspace{2mm}
  \\
\les \={\mathbb{G}}^\sharp_{\,2}\,(\#q,\omega') \ris_{21} = \les -
\le \#q \. \hat{\#x} \ri \le \#q \. \hat{\#y} \ri
 - i \tilde{\kappa}^z_2 \le -  \tilde{\kappa}^g_2 + \#q\.\hat{\#z} \ri \ris \Lambda_2 \vspace{2mm}
\\
\les \={\mathbb{G}}^\sharp_{\,2}\,(\#q,\omega') \ris_{22} = \les -
\le \#q \. \hat{\#y} \ri^2+ \tilde{\kappa}^t_2 \tilde{\kappa}^z_2
\ris \Lambda_2 \vspace{2mm}
\\
 \les \={\mathbb{G}}^\sharp_{\,2}\,(\#q,\omega')
\ris_{23} = \les  \#q \. \hat{\#y} \le \#q \. \hat{\#z}  -
\tilde{\kappa}^g_2 \ri - i  \tilde{\kappa}^t_2 \#q \. \hat{\#x}
\ris \Lambda_2 \vspace{2mm}
\\
\les \={\mathbb{G}}^\sharp_{\,2}\,(\#q,\omega') \ris_{31} = \les
 \#q \. \hat{\#x} \le  \#q \. \hat{\#z} - \tilde{\kappa}^g_2 \ri -
  i \tilde{\kappa}^t_2  \#q \. \hat{\#y}  \ris \Lambda_2 \vspace{2mm}
  \\
  \les \={\mathbb{G}}^\sharp_{\,2}\,(\#q,\omega')
\ris_{32} = \les  \#q \. \hat{\#y} \le \#q \. \hat{\#z}  -
\tilde{\kappa}^g_2 \ri + i  \tilde{\kappa}^t_2 \#q \. \hat{\#x}
\ris \Lambda_2 \vspace{2mm}
\\
\les \={\mathbb{G}}^\sharp_{\,2}\,(\#q,\omega') \ris_{33} = \les
\le \tilde{\kappa}^t_2 - \tilde{\kappa}^g_2 + \#q \. \hat{\#z} \ri
\le \tilde{\kappa}^t_2 + \tilde{\kappa}^g_2 - \#q \. \hat{\#z} \ri
\ris \Lambda_2 \vspace{2mm}
\end{array}
\right\},
\end{equation}
with
\begin{eqnarray}
\frac{1}{\Lambda_2} &=& \frac{1}{2} \lec \les \le \#q\.\hat{\#z}
\ri^2 - \le \#q\.\hat{\#x} \ri^2 - \le \#q\.\hat{\#y} \ri^2 \ris
\le \tilde{\kappa}^z_2 - \tilde{\kappa}^t_2 \ri - \#q\.\#q \le
\tilde{\kappa}^t_2 + \tilde{\kappa}^z_2 \ri \ric \nonumber \\ && +
\le \tilde{\kappa}^g_2 \ri^2 + 2 \tilde{\kappa}^g_2
\tilde{\kappa}^z_2 \#q\. \hat{\#z} + \le \tilde{\kappa}^t_2
\ri^2\,. \l{DGF_b}
\end{eqnarray}
Having established the spectral DGFs
$\={\mathbb{G}}^\sharp_{\,1,2}\,(\#q,\omega')$, we obtain the $\Sigma$ field
phasors   generated by the source phasors $\#J_{\,e,m} (\#r, \omega)$ as
\begin{eqnarray}
\#E (\#r, \omega) &=& \#{\mathring{E}} (\#r, \omega) \nonumber \\
&& + \frac{1}{4 \pi^3}\int_{V} \Bigg( \lec \int_{\#q}
\={\mathbb{G}}^\sharp_{\,1}\,(\#q,\omega') \exp \les i \, \#q \.
\le \#r - \#s \ri \ris \; d \#q
 \ric\.
 \les i \eta' (\omega')\, \#J_{\,e} (\#r, \omega ) - \#J_{\,m}(
\#r, \omega)\,
 \ris\nonumber \\ &&
- i \eta' (\omega') \lec \int_{\#q}
\={\mathbb{G}}^\sharp_{\,2}\,(\#q,\omega') \exp \les i \, \#q \.
\le \#r - \#s \ri \ris \; d \#q
 \ric\.
 \les   \#J_{\,e} (\#r, \omega ) - \frac{i}{\eta' (\omega')}\, \#J_{\,m}(
\#r, \omega)\,
 \ris
  \Bigg) \,d \#s \nonumber \\ &&
\end{eqnarray}
and
\begin{eqnarray}
\#H (\#r, \omega) &=& \#{\mathring{H}} (\#r, \omega) \nonumber
\\ && + \frac{1}{4 \pi^3}\int_{V} \Bigg( \lec \int_{\#q}
\={\mathbb{G}}^\sharp_{\,2}\,(\#q,\omega') \exp \les i \, \#q \.
\le \#r - \#s \ri \ris \; d \#q
 \ric\.
 \les  \#J_{\,e} (\#r, \omega ) - \frac{i}{ \eta' (\omega')}\,\#J_{\,m}(
\#r, \omega)\,
 \ris\nonumber \\ &&
- \frac{i}{ \eta' (\omega') }\lec \int_{\#q}
\={\mathbb{G}}^\sharp_{\,1}\,(\#q,\omega') \exp \les i \, \#q \.
\le \#r - \#s \ri \ris \; d \#q
 \ric\.
 \les  i \eta' (\omega')\,  \#J_{\,e} (\#r, \omega ) -  \#J_{\,m}(
\#r, \omega)\,
 \ris
  \Bigg) \,d \#s \,.\nonumber \\ &&
\end{eqnarray}

\section{Discussion}

 The Tellegen constitutive relations for an
isotropic chiral medium  moving at constant velocity are
presented in \r{cr1} and \r{cr2}.
 The availability of these constitutive relations facilitates a full analysis
of the planewave characteristics of the medium, and also enables
the spectral DGFs to be derived in a convenient form. In contrast
to earlier studies, the analysis presented herein is not
restricted to low relative speeds (Hillion 1993; Ben--Shimol \&
Censor 1995, 1997). Furthermore, the constitutive relations
\r{cr1} and \r{cr2} establish that the uniformly moving isotropic
chiral medium in fact belongs to the category of FCMs.

 In \S\ref{pw_section}, the
analysis of planewave propagation in reference frames $\Sigma'$
and $\Sigma$, is aided by the introduction of the Beltrami field
phasors $\#{\mathbb{Q}}'_{\,1,2}\,(\#r',\omega')$ in \r{Q12def}
and the Beltrami--like field phasors
$\#{\mathbb{Q}}_{\,1,2}\,(\#r,\omega)$  in \r{Q12def_x},
respectively. They facilitate a decoupling of the  Maxwell curl
postulates.
 A key property
of   $\#{\mathbb{Q}}'_{\,1,2}\,(\#r',\omega')$  is that
the  curl of $\#{\mathbb{Q}}'_{\,1,2}\,(\#r',\omega')$ is a
scalar multiple of $\#{\mathbb{Q}}'_{\,1,2}\,(\#r',\omega')$, as
demonstrated in \r{B12}. Such fields are known as Beltrami fields
and their properties are firmly established (Hillion \&
Lakhtakia 1993; Lakhtakia 1994b,b). In
contrast,
 the curl of
$\#{\mathbb{Q}}_{\,1,2}\,(\#r,\omega)$
 is not generally  parallel to
$\#{\mathbb{Q}}_{\,1,2}\,(\#r,\omega)$, as may be observed from \r{BB12}.  The
 Beltrami--like field phasors $\#{\mathbb{Q}}_{\,1,2}\,(\#r,\omega)$ therefore represent
 an important
extension of the usual Beltrami field concept which can be traced back to at least
the late 1880s (Beltrami 1889; Silberstein 1907; Trkal 1919).

Owing to their relatively large parameter space, linear bianisotropic
mediums support a richer palette of planewave properties than do
anisotropic and isotropic mediums, as has been highlighted lately
by investigations of NPV propagation (Mackay \& Lakhtakia 2004b)
and optical singularities (Berry 2005).  The planewave study
presented in \S\ref{pw_section} reveals that the bianisotropic
FCM described by the constitutive relations \r{cr2} generally
supports four independent wavenumbers for each direction of
propagation,
 from the perspective of a non--co--moving observer (the exception being propagation perpendicular
to the direction of translation for which only two independent
wavenumbers are supported). This constrasts with the two independent
wavenumbers supported by the isotropic chiral medium from the
perspective of the co--moving observer. We see in Figure~\ref{fig3}
that an isotropic chiral medium, which does not support NPV
propagation from the perspective of a co--moving observer, does
support NPV propagation from the perspective of a certain class of
non--co--moving observers, provided that the relative speed is
sufficiently high. This finding is consistent with results presented
for an  isotropic dielectric--magnetic medium moving at constant
velocity (Mackay \& Lakhtakia 2004a).   This is also consistent with
a study which showed that a FCM arising as a homogenized composite
medium can support NPV propagation provided that the gyrotropy
parameter of the gyrotropic constituent medium is sufficiently large
(Mackay \& Lakhtakia 2004b).

While explicit representations of DGFs are available for isotropic
mediums, these  are generally not available for  anisotropic and
bianisotropic mediums (Mackay \& Lakhtakia 2006). However, as
shown in \S\ref{DGF_section}, the field phasors for the FCM
described by  the constitutive relations  \r{cr2} may be
formulated in terms of spectral DGFs. By exploiting the
constitutive relations \r{cr2} and the Beltrami--like field
phasors $\#{\mathbb{Q}}_{\,1,2}\,(\#r,\omega)$, a convenient
representation of the spectral DGFs is established in
\r{DGF_a}--\r{DGF_b}.

\vspace{10mm}

\noindent{\bf Acknowledgement:} TGM acknowledges EPSRC for support
under grant GR/S60631/01.

\section*{Appendix 1}

As indicated by \r{w_NPV}, whether or not plane waves propagate
with negative phase velocity
 in a Faraday chiral medium is determined by the sign of the
parameter $\Omega (\tilde{k})$. The form  of $\Omega (\tilde{k})$
is provided here; a complete description of the derivation of
$\Omega (\tilde{k})$ is available elsewhere (Mackay \& Lakhtakia
2004a).

As a function of the relative wavenumber $\tilde{k}$, $\Omega
(\tilde{k})$ may be expressed in terms of the constitutive
parameters \r{e_cr}--\r{m_cr} and the wavevector orientation angle
$\theta$ as
\begin{eqnarray}
\Omega (\tilde{k}) &=& \mbox{Re} \lec \tilde{k} \ric \times
\mbox{Re} \, \Bigg\{
 \,\frac{1}{\mu^{z*} (\omega')}
 \les \tilde{k}^* \sin \theta  -
i  \tau^* \xi^{z*} (\omega')  \ris \,\sin \theta \nonumber \\ && +
\frac{1}{ \les \mu^{t*} (\omega') \ris^2  - \les \mu^{g*} (\omega')
\ris^2} \, \Bigg[ \tilde{k}^* \Bigg(
 \les
\mu^{t*} (\omega') \le |\alpha|^2 +1 \ri + i \mu^{g*} (\omega') \le
\alpha - \alpha^* \ri \ris \, \cos^2 \theta  \nonumber \\ &&  + |
\tau |^2 \mu^{t*} (\omega')  \sin^2 \theta - \les \mu^{t*} (\omega')
\le \alpha^* \tau + \alpha \tau^* \ri + i \mu^{g*} (\omega') \le
\tau - \tau^* \ri \ris \, \sin \theta \cos \theta \Bigg) \nonumber
\\ && + \les \mu^{t*} (\omega') \xi^{g*} (\omega') - \mu^{g*} (\omega')
\xi^{t*} (\omega') \ris \les \le |\alpha|^2 +  1 \ri \, \cos \theta
- \alpha^* \tau \, \sin \theta \ris \nonumber
\\ && - i \les \mu^{t*} (\omega') \xi^{t*} (\omega') - \mu^{g*} (\omega') \xi^{g*}
 (\omega') \ris \les \le \alpha -
\alpha^* \ri\,  \cos \theta - \tau \,  \sin \theta \ris \Bigg]
\Bigg\}
 \,. \l{w_xz}
\end{eqnarray}
Herein,
\begin{equation}
\left.
\begin{array}{l}
\alpha = \displaystyle{ \frac{ L_{12} L_{33} + L_{13} L_{23} } {
L^2_{13}
-  L_{11}  L_{33}} }\\
\vspace{-3mm}
\\
\tau = \displaystyle{ \frac{  L_{12} L_{23} - L_{13} L_{22} } {
L_{13}  L_{23} +
  L_{12}   L_{33} }}
\end{array}
\right\}, \l{alp_bet}
\end{equation}
with
\begin{equation}
\left. \begin{array}{l} L_{11}= \displaystyle{\eps^t(\omega') +
\frac{2 \Gamma \mu^g (\omega')\, \xi^t (\omega') - \mu^t(\omega')
\lec \les \xi^{t} (\omega') \ris^2 + \Gamma^2 \ric}{\les
\mu^t(\omega') \ris^2 - \les \mu^g(\omega') \ris^2}} \vspace{2mm}
\\
L_{22} = \displaystyle{\eps^t (\omega') - \frac{\tilde{k}^2 \sin^2
\theta}{\mu^z (\omega') } + \frac{ 2 \Gamma  \mu^g (\omega') \,
\xi^t (\omega') - \mu^t (\omega') \lec \les  \xi (\omega') \ris^2
+ \Gamma^2 \ric }{\les \mu^t (\omega')\ris^2
- \les \mu^g (\omega') \ris^2}} \vspace{2mm}\\
L_{33} = \displaystyle{\eps^z (\omega') - \frac{\les \xi^z
(\omega') \ris^2}{\mu^z (\omega')} - \frac{\mu^t (\omega') \,
\tilde{k}^2 \sin^2 \theta}{\les \mu^t (\omega') \ris^2
- \les \mu^g (\omega') \ris^2}} \vspace{2mm}\\
L_{12} =\displaystyle{ i \le \eps^g (\omega') + \frac{\mu^g
(\omega') \lec \les \xi^t (\omega') \ris^2 + \Gamma^2 \ric - 2
\Gamma  \mu^t (\omega') \,\xi^t (\omega')}{ \les \mu^t
(\omega')\ris^2
- \les  \mu^g (\omega') \ris^2} \ri} \vspace{2mm}\\
L_{13} = \displaystyle{ \frac{ \Gamma  \mu^t (\omega') - \mu^g
(\omega') \xi^t (\omega') }{\les \mu^t (\omega') \ris^2
- \les \mu^g (\omega') \ris^2} \, \tilde{k} \, \sin \theta} \vspace{2mm} \\
L_{23} = \displaystyle{ i  \les \frac{\Gamma \mu^g (\omega') -
\mu^t (\omega') \xi^t (\omega') }{\les \mu^t (\omega') \ris^2 -
\les \mu^g (\omega') \ris^2}- \frac{\xi^z (\omega') }{\mu^z
(\omega')} \ris \, \tilde{k} \, \sin \theta} \end{array} \right\}
\end{equation}
and $\Gamma = \xi^g (\omega') + \tilde{k} \, \cos \theta $.

\newpage

\newpage

\begin{figure}[!ht]
\centering
\psfull\epsfig{file=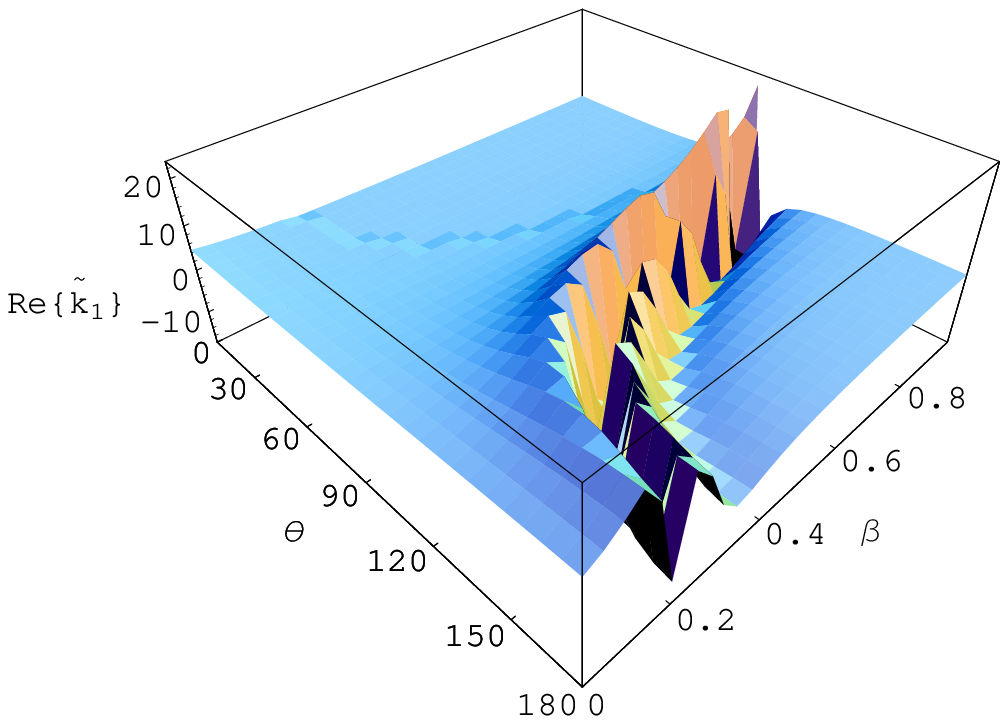,width=2.8in}\hspace{10mm}
\epsfig{file=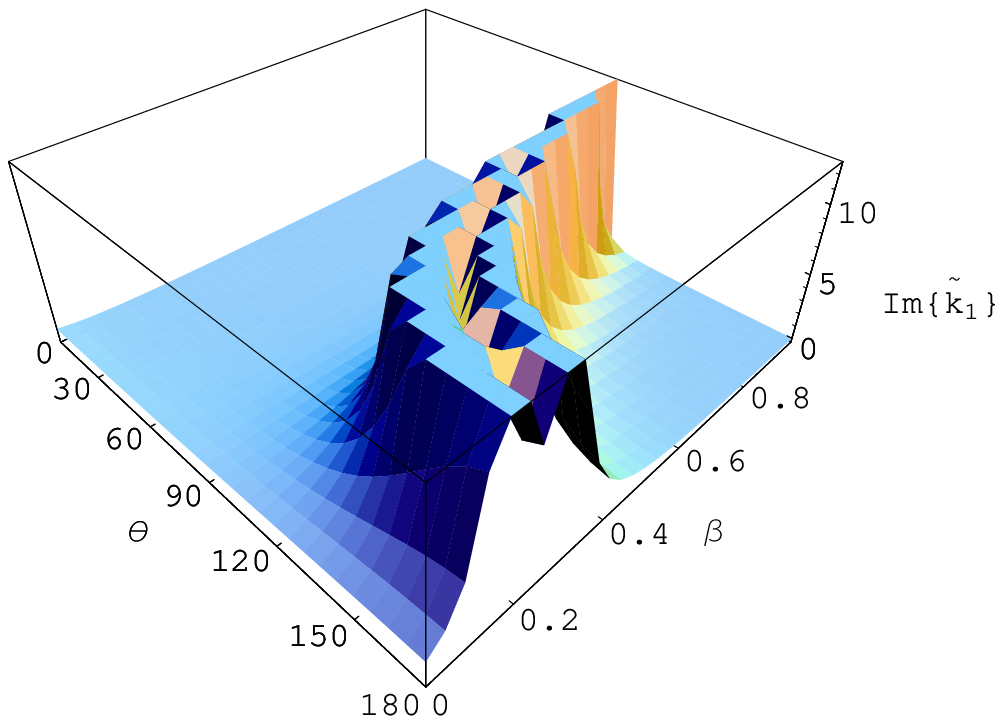,width=2.8in}\\ \vspace{10mm}
\epsfig{file=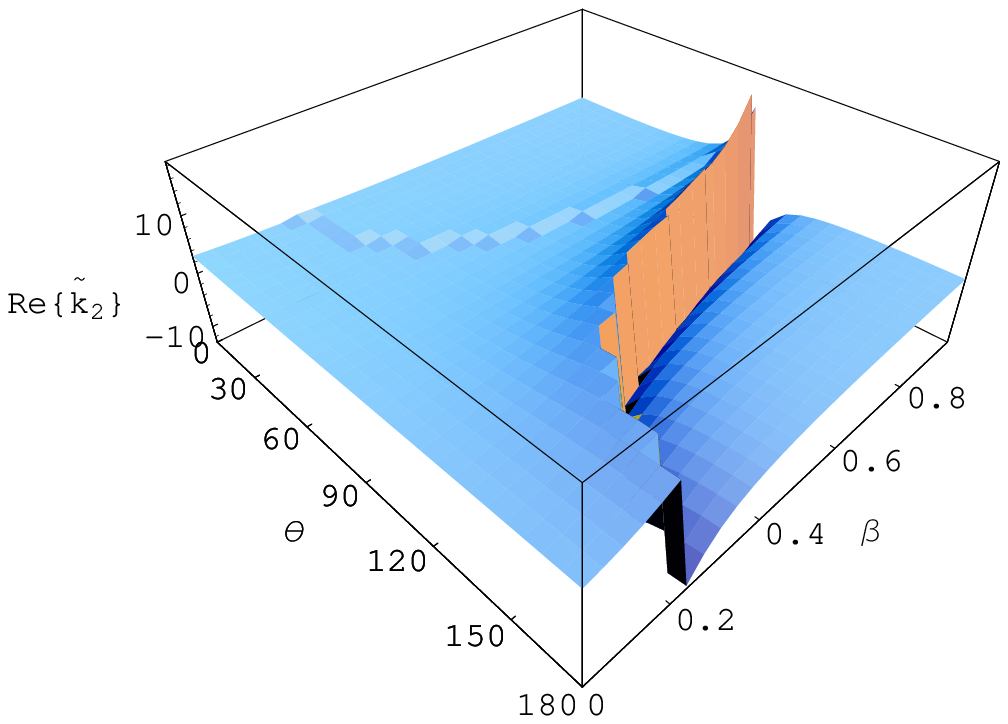,width=2.8in}\hspace{10mm}
\epsfig{file=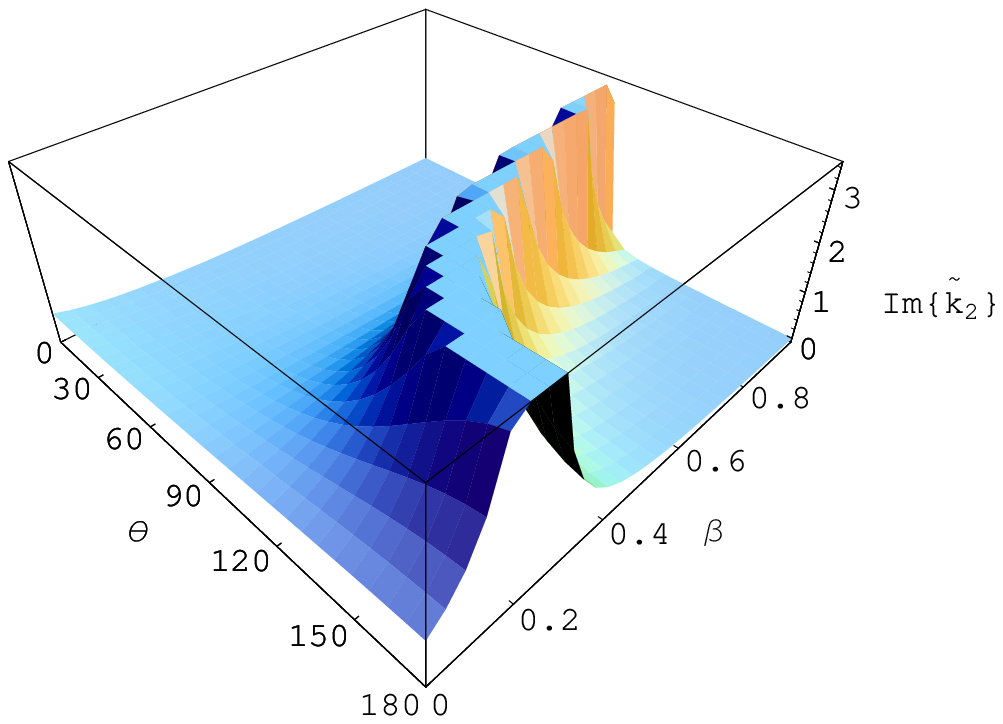,width=2.8in}
  \caption{\label{fig2} Real (left) and imaginary (right) parts of the
  four relative wavenumbers $\tilde{k}_{1,2,3,4}$  plotted against the
relative speed $\beta = v/\co$ and the propagation  angle $\theta =
\cos^{-1} \hat{\#v}\.\hat{\#k}$ (in degree). Constitutive parameters
in $\Sigma'$: $\eps'_{\mbox{\tiny{}}} =  6.5 + i 1.5$,
$\xi'_{\mbox{\tiny{}}} = 1 + i 0.2 $, and $\mu'_{\mbox{\tiny{}}} =
3.0 + i 0.5$.}
\end{figure}

\newpage

\setcounter{figure}{1}
\begin{figure}[!ht]
\centering \psfull
\epsfig{file=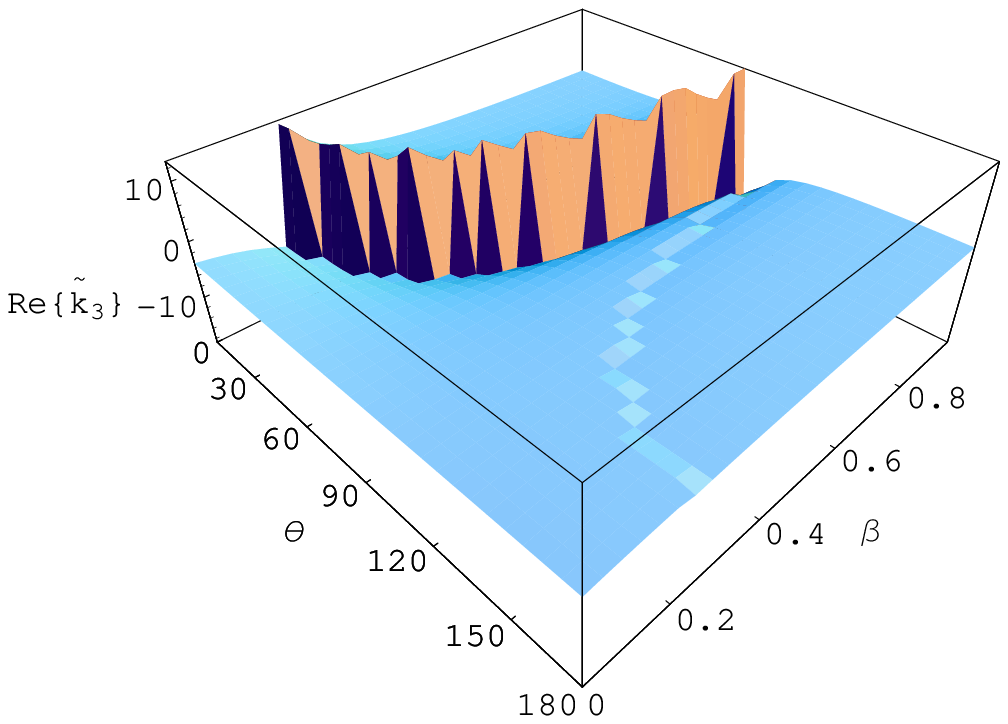,width=2.8in}\hspace{10mm}
\epsfig{file=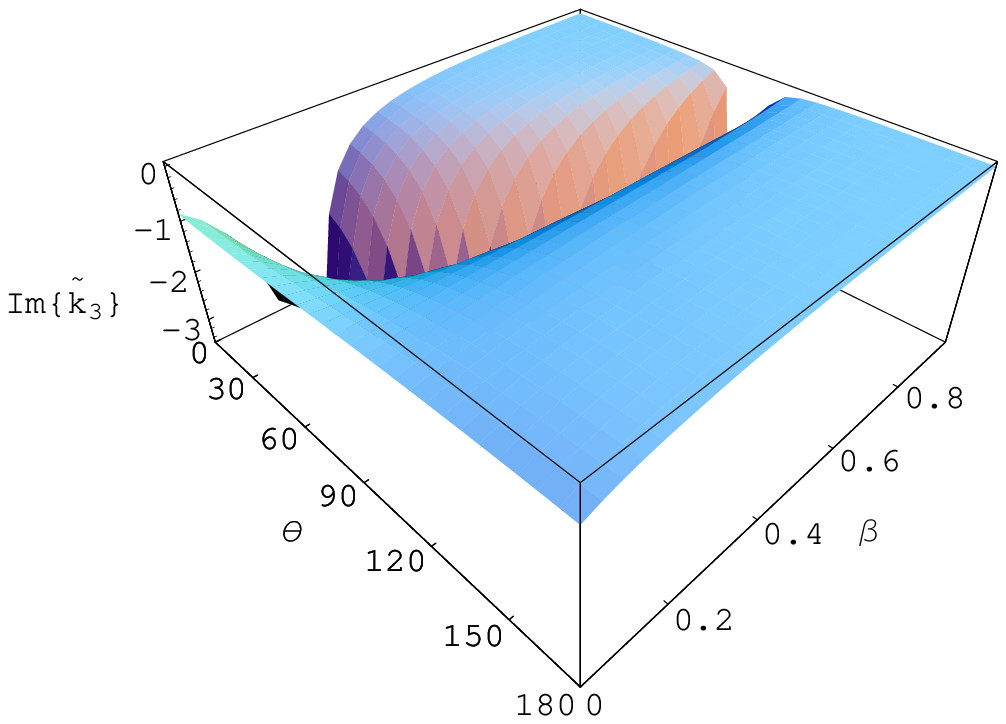,width=2.8in}\\\vspace{10mm}
\epsfig{file=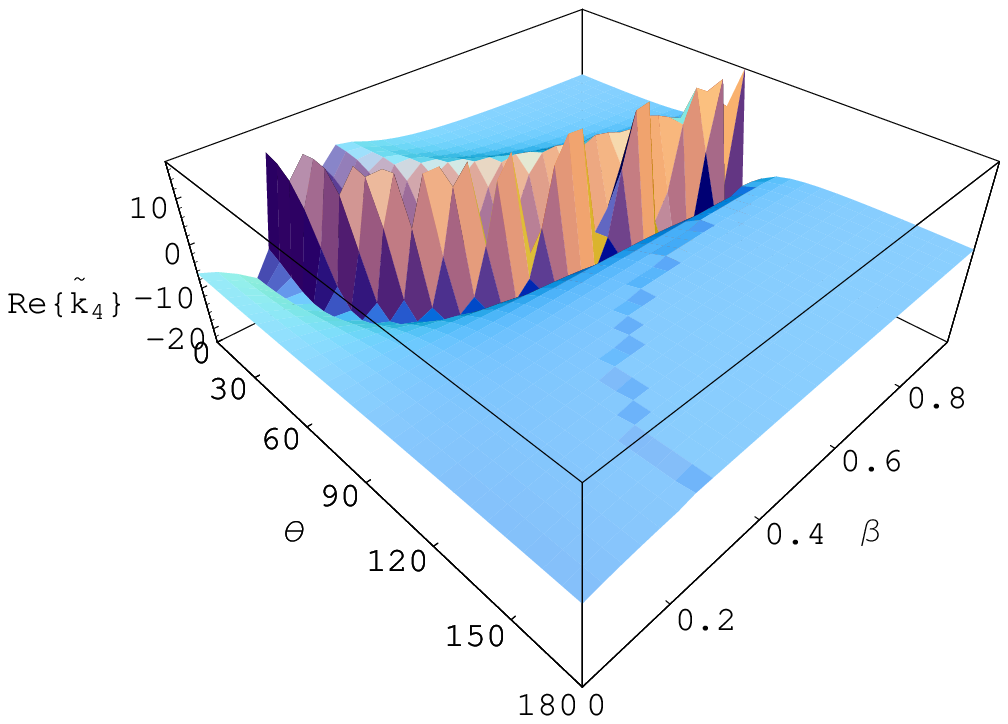,width=2.8in}\hspace{10mm}
\epsfig{file=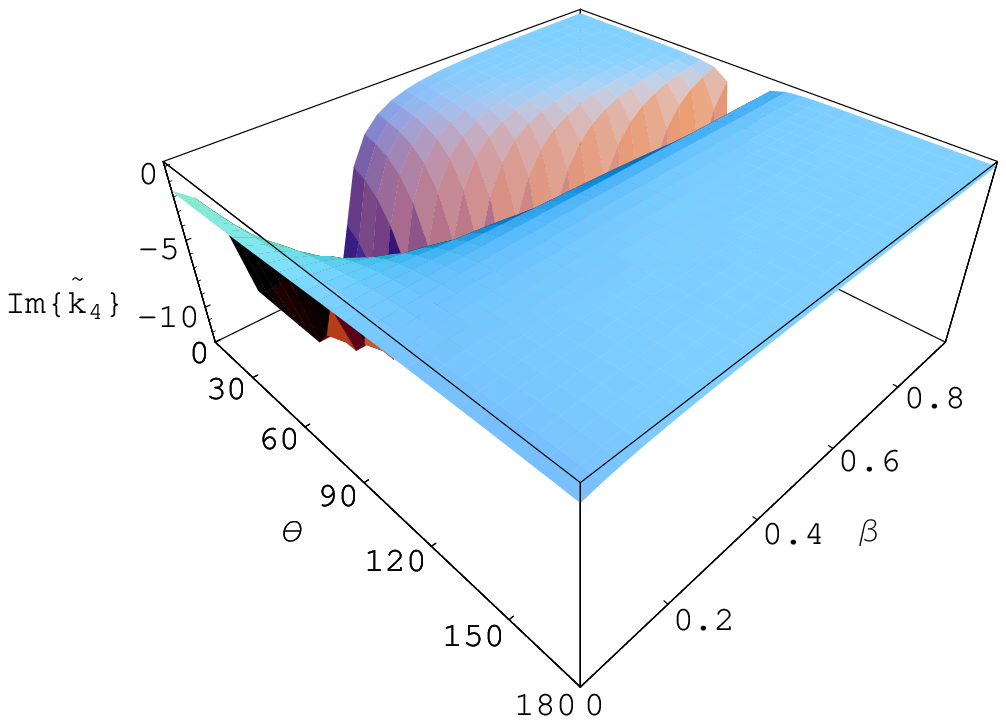,width=2.8in}
  \caption{continued}
\end{figure}

\newpage
\begin{figure}[!ht]
\centering \psfull
\epsfig{file=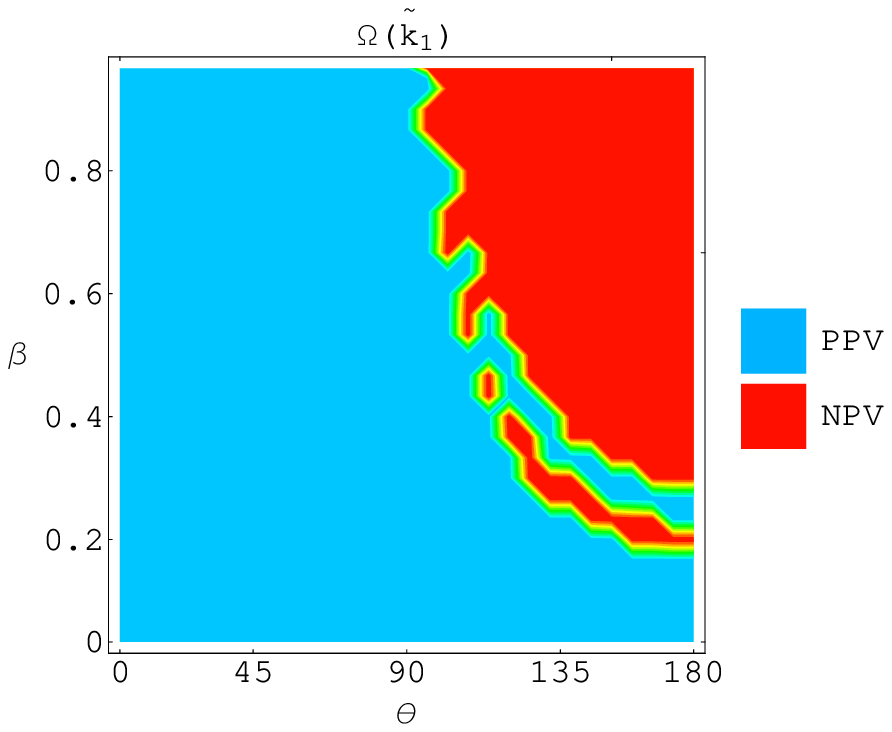,width=2.8in}\hspace{10mm}
\epsfig{file=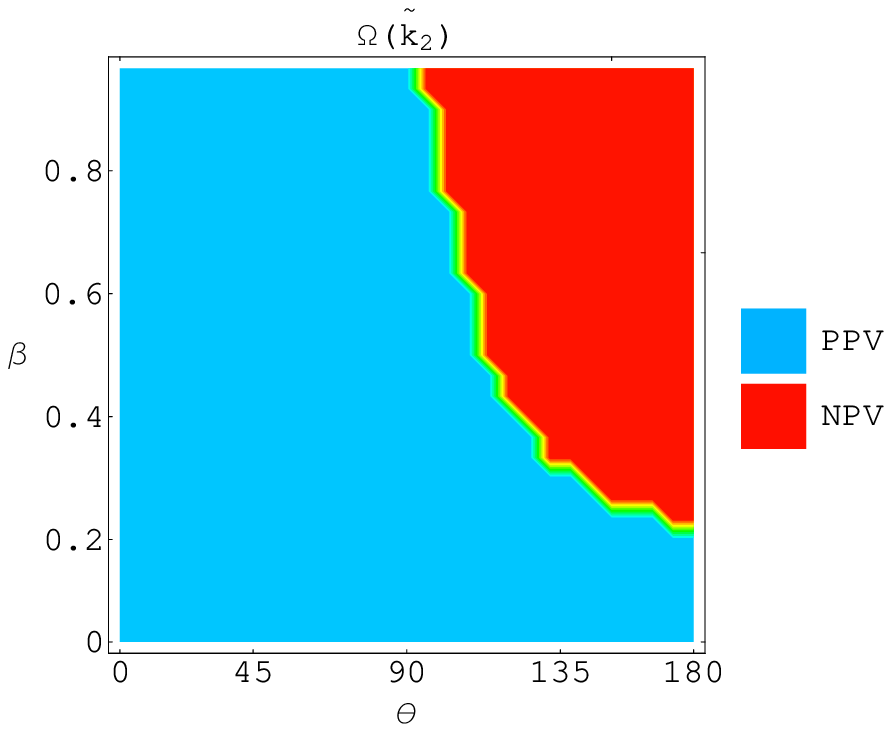,width=2.8in}\\\vspace{10mm}
\epsfig{file=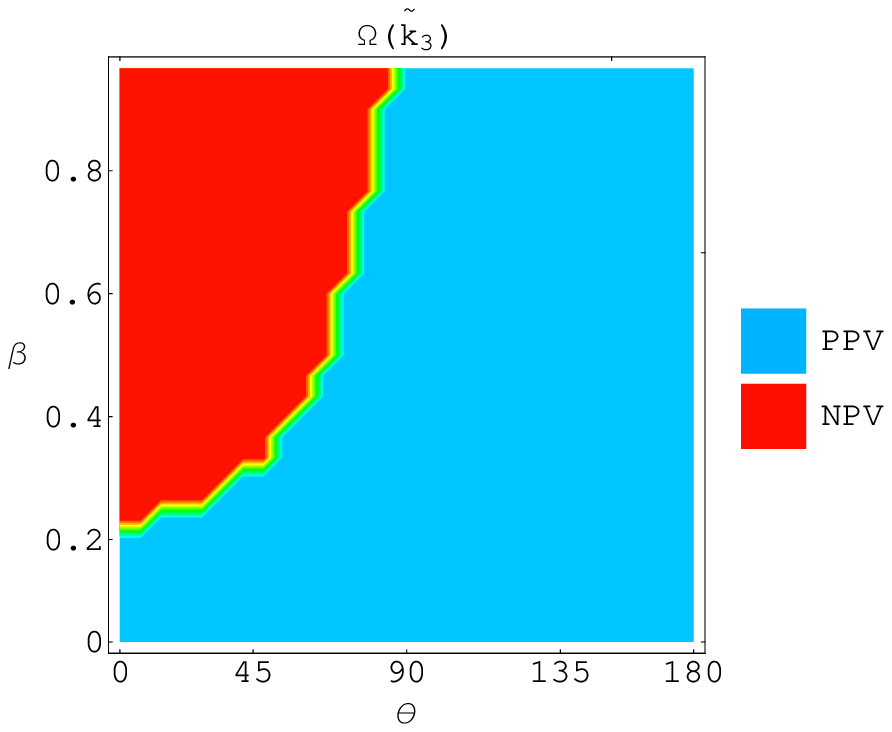,width=2.8in}\hspace{10mm}
\epsfig{file=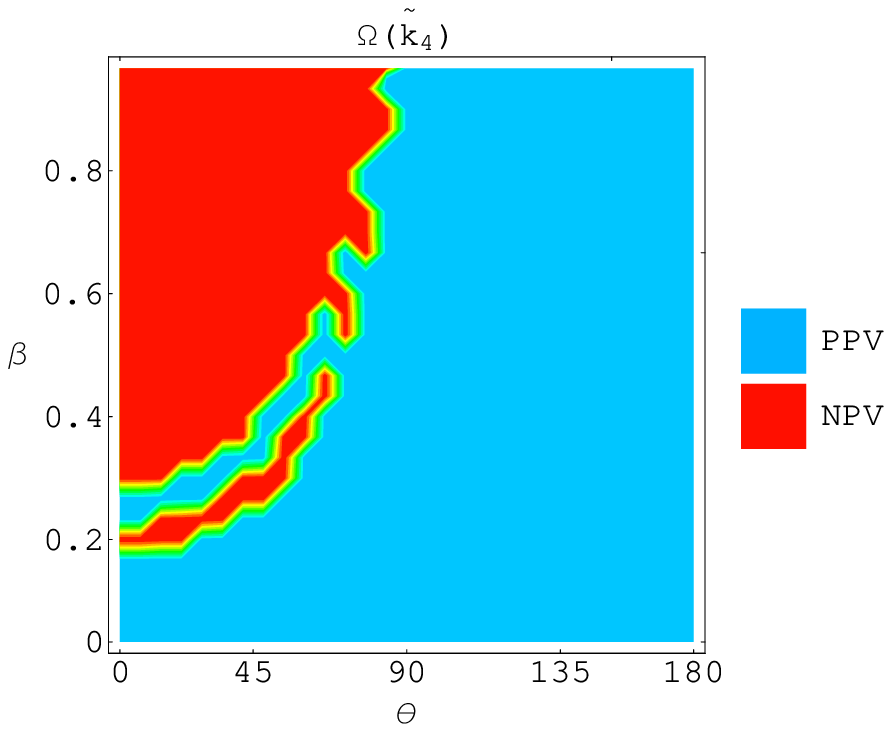,width=2.8in}
  \caption{\label{fig3} $\beta\theta$--regimes of NPV and PPV,
as  viewed from the inertial frame $\Sigma$, with
  respect to the relative speed $\beta = v/\co$ and
  the propagation  angle $\theta =
\cos^{-1} \hat{\#v}\.\hat{\#k}$ (in degree), as determined by the
parameter $\Omega(\tilde{k})$ for the
  four relative wavenumbers $\tilde{k} \in \lec \tilde{k}_{1}, \tilde{k}_{2}, \tilde{k}_{3}, \tilde{k}_{4} \ric$.
 Constitutive
parameters in $\Sigma'$: $\eps'_{\mbox{\tiny{}}} =  6.5 + i 1.5$,
$\xi'_{\mbox{\tiny{}}} = 1 + i 0.2 $, and $\mu'_{\mbox{\tiny{}}} =
3.0 + i 0.5$. }
\end{figure}


\begin{thebibliography}{99}

\bibitem{Beltrami}
Beltrami, E. 1889 Considerazioni idrodinamiche. \emph{Rend. Inst. Lombardo Acad. Sci. Lett.} {\bf 22}, 122--131.
 An English translation of the paper by G. Filipponi is available: Beltrami, E. 1985 Considerations on hydrodynamics. \emph{Int. J. Fusion Energy}
{\bf 3}(3), 53--57.

\bibitem{Berry}
Berry, M.V. 2005 The optical singularities of bianisotropic
crystals. \emph{Proc. R. Soc. Lond. A} {\bf 461}, 2071--2098.

\bibitem{Censor95}
Ben--Shimol, Y. \&  Censor,  D. 1995 Wave propagation in moving chiral
media: Fizeau's experiment revisited. \emph{Radio Sci.} {\bf 30},
1313--1324.

\bibitem{Censor97}
Ben--Shimol, Y. \&  Censor,  D. 1997 First order propagation in moving
chiral media. \emph{Radio Sci.} {\bf 32},  2201--2207.

\bibitem{Bohren}
Bohren, C.F. 1974 Light scattering by an optically active sphere.
\emph{Chem. Phys. Lett.} {\bf 29}, 459--462.

\bibitem{Ceperley}
Ceperley, P.H. 1992  Rotating waves. \emph{Am. J. Phys.} {\bf 60},
938--942.

\bibitem{Chan-1}
Chandrasekhar, S. 1956 On force--free magnetic fields. \emph{Proc.
Natl. Acad. Sci. USA} {\bf 42}, 1--5.

\bibitem{Chan-2}
Chandrasekhar, S. 1957 On cosmic magnetic fields. \emph{Proc. Natl.
Acad. Sci. USA} {\bf 43}, 24--27.

\bibitem{Chawla}
Chawla, B.R. \&  Unz, H. 1971 {\em Electromagnetic waves in moving
magneto--plasmas\/}. Lawrence, KS, USA: University Press of
Kansas.

\bibitem{Chen}
Chen, H.C. 1983 {\em Theory of electromagnetic waves\/}. New York,
NY, USA: McGraw--Hill.


\bibitem{Dritschel}
Dritschel, D.G. 1991 Generalized helical Beltrami flows in
hydrodynamics and magnetohydrodynamics. \emph{J. Fluid Mech.} {\bf
222}, 525--541.


\bibitem{Engheta92}
Engheta, N, Jaggard,  D.L. \&   Kowarz, M.W. 1992 Electromagnetic waves
in Faraday chiral media. \emph{IEEE Trans. Antennas Propagat.} {\bf
40},  367--374.


\bibitem{Engheta89}
Engheta, N., Kowarz,  M.W. \&   Jaggard,  D.L.  1989 Effect of chirality on
the Doppler shift and aberration of light waves. \emph{J. Appl.
Phys.}  {\bf66},  2274--2277.


\bibitem{Hillion}
Hillion, P. 1993 Electromagnetism in a moving chiral medium.
\emph{Phys. Rev. E} {\bf 48}, 3060--3065.

\bibitem{HL93}
Hillion, P. \& Lakhtakia, A. 1993 On an initial boundary value
problem involving Beltrami--Moses fields in electromagnetic
theory. \emph{Phil. Trans. R. Soc. Lond. A} {\bf 344}, 235--248;
corrections: 1994 {\bf 347}, 543.



\bibitem{Hinders}
Hinders, M.K., Trott,  K.D., Moses,  H.E., Nagem, R.J., Konstantopoulos, D.,
Rhodes,  B.A. \&  Sandri, G.vH.  1991 Transmission through a moving
chiral slab.  \emph{J. Opt. Soc. Am. B} {\bf 8}, 1958--1961.


\bibitem{Kong}
 Kong, J.A. 1986 \emph{Electromagnetic wave theory}. New York,
NY, USA:  Wiley.

\bibitem{Krowne}
 Krowne,  C.M. 1984  Electromagnetic theorems for complex anisotropic
media. \emph{
 IEEE Trans. Antennas Propagat.} {\bf 32},
1224--1230.


\bibitem{Beltrami-1}
Lakhtakia, A. 1994a   Viktor Trkal, Beltrami fields, and Trkalian flows.
\emph{Czechoslovak J. Phys.} {\bf 44}, 89--96.


\bibitem{Beltrami-2}
Lakhtakia, A. 1994b   \emph{Beltrami fields in chiral media}.
Singapore: World Scientific.


\bibitem{LMW03}
Lakhtakia,  A.,  McCall,  M.W. \&   Weiglhofer,  W.S.  2003 Negative
phase--velocity mediums. In
  \emph{Introduction to complex mediums for optics and
electromagnetics},  eds.  W.S. Weiglhofer \&  A. Lakhtakia.
Bellingham, WA, USA: SPIE  Press. pp. 347--363.

\bibitem{LVV91}
Lakhtakia, A., Varadan,  V.V. \&   Varadan, V.K.  1991 Plane wave scattering
response of a simply moving electrically small, chiral sphere.
\emph{J. Mod. Opt.} {\bf 38},  1841--1847.



\bibitem{LW96}
Lakhtakia,  A. \&   Weiglhofer,  W.S. 1996 Lorentz covariance, Occam's
razor, and a constraint on linear constitutive relations. \emph{
Phys. Lett.  A } {\bf 213}, 107--111; correction 1996 {\bf 222}, 459.

\bibitem{Lax}
Lax,  B. \&  Button,  K.J. 1962 \emph{Microwave ferrites and
ferrimagnetics}.  New York, NY, USA: McGraw--Hill.



\bibitem{M05}
Mackay,  T.G.  2005 Plane waves with negative phase velocity in
isotropic chiral mediums. \emph{Microwave Opt. Technol. Lett.} {\bf
45}, 120--121; erratum 2005 {\bf 47}, 406.



\bibitem{ML04}
Mackay, T.G. \&   Lakhtakia, A. 2004a  Negative phase velocity in
a uniformly moving, homogeneous, isotropic, dielectric--magnetic
medium. \emph{J. Phys. A: Math. Gen.} {\bf 37}, 5697--5711.

\bibitem{ML_PRE}
Mackay.  T.G.  \&  Lakhtakia, A. 2004b Plane waves with negative phase
velocity in Faraday chiral mediums. \emph{Phys. Rev. E} {\bf 69},
026602.


\bibitem{ML_PiO}
Mackay.  T.G.  \&  Lakhtakia, A. 2006 Electromagnetic fields in
linear bianisotropic mediums. \emph{Prog. Optics} (to appear).

\bibitem{MLS06}
 Mackay,  T.G.,  Lakhtakia, A. \&  Setiawan, S. 2006
 Positive--, negative--, and orthogonal--phase--velocity
propagation of electromagnetic plane waves in a simply moving
medium.  \emph{Optik}  (accepted for publication)
(doi:10.1016/j.ijleo.2006.02.006).

\bibitem{Marc}
Marcinkowski, M.J. 1992 The underlying unity between
magnetostatics, fluid flow and deformation of solids. \emph{Acta
Phys. Polo. A} {\bf 81}, 543--558.

\bibitem{McL}
McLaughlin, D. \& Pironneau, O. 1991 Some notes on periodic
Beltrami fields in Cartesian geometry. \emph{J. Math. Phys.} {\bf
32}, 797--804.

\bibitem{Melrose}
Melrose, D.B. \& McPhedran, R.C. 1991 \emph{Electromagnetic
processes in dispersive media}. Cambridge, UK: Cambridge University
Press.

\bibitem{SAR}
Ramakrishna, S.A. 2005 Physics of negative refractive index materials.
\emph{Rep. Progr. Phys.} {\bf 68}, 449--521.

\bibitem{Pappas}
Pappas, C.H. 1965 {\em Theory of electromagnetic wave
propagation\/}. New York, NY, USA: McGraw--Hill.


\bibitem{Silberstein}
Silberstein, L. 1907 Elektromagnetische Grundgleichungen in Bivektorieller Behandlung. \emph{Ann. Phys. Lpz.} {\bf 22}, 579--587.



\bibitem{Tai}
 Tai, C.T. 1994  \emph{Dyadic Green functions in electromagnetic
 theory}
2nd ed. Piscataway, NJ, USA: IEEE Press.

\bibitem{Trkal}
Trkal, V. 1919 Pazn\'amka hydrodynamice vazk\'ych tekutin.
\emph{\u{C}asopis pro P\u{e}stov\'an\'i Mathematiky a Fysiky}
{\bf 48}, 302--311. An English translation by I. Gregora is available:
Trkal V. 1994 A note on the hydrodynamics of viscous fluids. \emph{Czechoslovak J. Phys.} {\bf 44}, 97--106.

\bibitem{Walker}
Walker, J. S. 1988  \emph{Fourier analysis\/}.
New York, NY, USA: Oxford University Press.

\bibitem{WL98}
Weiglhofer,  W.S. \&   Lakhtakia,  A. 1998 The correct constitutive
relations of chiroplasmas and chiroferrites. \emph{Microwave Opt.
Technol. Lett. } {\bf 17},  405--408.




\bibitem{WLM98}
Weiglhofer,  W.S.,   Lakhtakia,  A.  \&  Michel, B. 1998
On the constitutive parameters of
a chiroferrite composite medium.
\emph{Microwave Opt. Technol. Lett.} {\bf 18}, 342--345.

\bibitem{WM00}
Weiglhofer,  W.S.  \&  Mackay,  T.G. 2000
Numerical studies of the constitutive
parameters of a chiroplasma composite medium.
\emph{Arch.  Elektr. \"{U}bertrag.} {\bf 54}, 259--265.

\bibitem{Yoshida}
Yoshida, Z. 1991  Helicity waves propagating in a plasma. \emph{J.
Plasma Phys.} {\bf 45}, 481--488.




\end{thebibliography}
\end{document}